\documentclass{article}

\usepackage{arxiv}

\usepackage[T1]{fontenc}    
\usepackage[colorlinks,citecolor=blue,linkcolor=blue]{hyperref}
\usepackage{hyperref}       
\usepackage{url}            
\usepackage{booktabs}       
\usepackage{amsfonts}       
\usepackage{nicefrac}       
\usepackage{microtype}      
\usepackage{comment}
\usepackage{amsmath,amssymb,amsfonts} 
\usepackage{color,graphicx}                 
\usepackage{caption}
\graphicspath{{images/}} 
\usepackage{color}                    
\usepackage{nccmath}
\usepackage{mathrsfs}
\usepackage{tikz-cd}
\usepackage{cleveref}
\usetikzlibrary{positioning}
\usetikzlibrary{arrows}

\usepackage{enumerate}

\usepackage{ntheorem}
\DeclareGraphicsExtensions{.pdf}

\newtheorem{theorem}{Theorem}[section]

\newtheorem{remark}[theorem]{Remark}
\newtheorem{definition}[theorem]{Definition}
\newtheorem{exmp}{Example}[section]
\newtheorem{result}{Result}[section]
\newtheorem*{proof-non}{\underline{Proof}}
\newtheorem*{solo-non}{\underline{Solution}}
\newtheorem*{verif-non}{\underline{Verification}}
\newtheorem*{hint-non}{\underline{Hint}}
\newtheorem*{note-non}{\underline{Note}}
\makeatletter
\renewcommand{\slimits@}{\limits}
\renewcommand{\nmlimits@}{\limits}
\makeatother
\newcommand{\bigslant}[2]{{\raisebox{.2em}{$#1$}\left/\raisebox{-.2em}{$#2$}\right.}}
\crefname{equation}{relations}{relations}
\usepackage{blindtext}
\usepackage{fancyhdr}
\usepackage{float}

\title{Non-commutative first order differential calculus over finitely generated associative algebras}

\date{September 10, 2019}	

\author{
  \textbf{Ali-Reza~Assar}\\ Vienna, Austria\\
  \texttt{assar.mehrdad@gmail.com} \\
\And
 \textbf{Roya~ Famili}\\ Razi University, Kermanshah, Iran\\
  \texttt{r.family285@gmail.com} \\
}

\begin{document}
\maketitle

\begin{abstract}
In this review article the construction of first order coordinate differential calculi on finitely generated and finitely related associative algebras are considered and explicit construction of the bimodule of one form over such algebras is presented. The concept of optimal algebras for such caculi are also discussed. Detailed computations presented will make this note particularly useful for physicists. 
\end{abstract}


\section{Introduction}
let $A$ be an associative algebra over a field $\mathbb{F}$. For our purpose, we need to work with a presentation of $A$ by generators and relations. so, let $\mathcal{F}$ be the free algebra on the generators $x^{1}, \ldots, x^{n}$, over a field $\mathbb{F}$, so
\begin{equation*}
\mathcal{F} = \mathbb{F}\langle x^{1}, \ldots ,x^{n}\rangle .
\end{equation*}
Let, $I= (f_{1}(x^{1}, \ldots, x^{n}), \ldots , f_{p}(x^{1},\ldots , x^{n}))\lhd \mathcal{F}$ be a 2-sided ideal in $\mathcal{F}$, and let
\begin{equation}
A:= \bigslant{\mathcal{F}}{I} = \bigslant{\mathbb{F}\langle x^{1}, \ldots , x^{n}\rangle}{\left( f_{1}(x^{1}, \ldots ,x^{n} ),\ldots , f_{p}(x^{1}, \ldots , x^{n})\right)}.
\end{equation}
and let
\begin{equation*}
\eta_{_{I}}: \left\lbrace\begin{array}{rl}\mathcal{F}&\longrightarrow A = \bigslant{\mathcal{F}}{I}\\
x^{i} &\longmapsto \bar{x}^{i}
\end{array}\right.
\end{equation*}
be the canonical algebra epimorphism, $Ker(\eta_{_{I}}) = I$ then
\begin{equation}
A \cong \mathbb{F}\langle \bar{x}^{1}, \ldots , \bar{x}^{n}\vert f_{i}(\bar{x}^{1}, \ldots , \bar{x}^{n}) = 0,\; i = 1, \ldots , p\rangle .
\end{equation}
and this gives a presentation of the algebra $A$ by means of generators and relations.\\
Let $M$ ba an $A$-bimodule such that $M$ is a free right A-module. Let $d: A \longrightarrow M$ be a derivation of $A$ into $M$, i.e. a linear mapping for which the Leibniz rule holds
\begin{equation*}
\forall f, g \in A:\quad d(fg) = (df).g +f.(dg),
\end{equation*}
Let us denote the $Im(d)$ in $M$ by $d(A)$. Let $\Omega_{d}^{1}(A)$ be the submodule of $M$ generated by $d(A)$, i.e.
\begin{equation}
\Omega_{d}^{1}(A) = A\cdot d(A)\cdot A = lin.span\lbrace f\cdot dg \cdot h\vert f, g, h\in A\rbrace . \label{q}
\end{equation}
Clearly $\Omega_{d}^{1}$ is also an $A$-bimodule that is free on the right, and 
\begin{equation}
d: A\longrightarrow \Omega_{d}^{1}(A).
\end{equation}
is a derivation of $A$ into $\Omega_{d}^{1}(A)$. Hence $\Omega_{d}^{1}(A)$ is an A-bimodule, generated by $dx^{1}, \ldots , dx^{n}\in M$ over $A$ such that it is free on the right. Notice that this means
\begin{equation*}
\Omega_{d}^{1}(A) = \sum_{i\oplus} dx^{i}\centerdot A\cong \underbrace{A\oplus A\oplus\cdots \oplus A}_{n-fold},\;as\;A-right\;module.
\end{equation*}
Notice that $f\cdot dg\cdot h = f\cdot d(gh)- (fg)\cdot dh$, for $f, g, h\in A$. Hence by the relation (\ref{q}), $\Omega_{d}^{1}(A) = A\cdot d(A) = d(A)\cdot A$. However the left and right action of $A$ in $\Omega_{d}^{1}(A)$ are not equal; i.e. $\Omega_{d}^{1}(A)$ is not taken to be a symmetric $A$-bimodule.\\
The pair $(\Omega_{d}^{1}(A), d)$ is called a first order differential calculus over $A$, FODC over $A$, for short. The derivation $d$ is said to be a free differential.\\
One must carefully note that in contrast to the classical case, the differential one forms $f\cdot dg$ and $dg\cdot f$ are not in general equal. The reason is not because $A$ may be noncommutative but rather it is because in general the A-A bimodule $\Omega^{1}(A)$ is not a symmetric bimodule! That is the left and the right actions of $A$ in $\Omega^{1}(A)$ are different. This is a feature that may well happen even when $A$ is a commutative algebra. We start with a few examples. 
\section{Examples }
\begin{exmp}\label{e}
Let $A = \mathbb{C}[x]$ be the polynomial algebra in one variable over the field of complex numbers $\mathbb{C}$. Let $\Omega^{1}(A)$ be the free right A-module generated by the element (symbol) $dx$ over $A$, i.e.
\begin{equation}
\Omega^{1}(A) := \lbrace dx\centerdot f\vert f\in A\rbrace .
\end{equation}
Fix a polynomial $p(x)\in \mathbb{C}[x]$. There exists a unique bimodule structure on $\Omega^{1}(A)$ such that the left action in $\Omega^{1}(A)$ is defined by
\begin{equation}
x\centerdot dx = dx\centerdot p(x).
\end{equation}
Let us denote this $A$-$A$ bimodule by $\Omega_{p}^{1}(A)$. It is easily seen that $\Omega_{p}^{1}(x)$ is a FODC over $A$ with the differential mapping defined by
\begin{equation}
d\left( \sum_{n}\alpha_{n}x^{n}\right) =\sum_{n}\sum_{i+j=n-1}\alpha_{n} x^{i}\centerdot dx\centerdot x^{j}. 
\end{equation}
Indeed we can write, using the linearity of $d$,
\begin{equation*}
d\left( \sum_{n}\alpha_{n}x^{n}\right) = \sum_{n}\alpha_{n}dx^{n}. \label{w}
\end{equation*}
and use the following steps
\begin{align*}
dx^{n} &= d(x\centerdot x^{n-1}) = dx\centerdot x^{n-1} + x\centerdot dx^{n-1}= dx\centerdot x^{n-1} + x\centerdot d(x\centerdot x^{n-2}),\\
&= dx\centerdot x^{n-1} + x\centerdot dx\centerdot x^{n-2}+ x^{2} \centerdot dx^{n-2}= \cdots = \sum_{i+j=n-1}x^{i}\centerdot dx\centerdot x^{j}.
\end{align*}
 from which the relation (\ref{w}) follows.
\end{exmp}
\begin{definition}
Two FODCalculi $(\Omega_{d}^{1}(A), d)$ and $(\Gamma_{d^{\prime}}^{1}(A), d^{\prime})$ are said to be isomorphic if there exists a bijective linear mapping (i.e. an isomorphism of vector spaces)
\begin{equation*}
\varphi : \Omega_{d}^{1}(A)\longrightarrow \Gamma_{d^{\prime}}^{1}(A).
\end{equation*}
such that the following diagram is commutative
\begin{equation*}
\begin{tikzpicture}
  \node (A) {$A$};
  \node (B) [below=of A] {$\Omega_{d}^{1}(A)$};
  \node (AS) [right=of A] {$A$};
  \node (D) [right=of B] {$\Gamma_{d^{\prime}}^{1}(A)$};
  \draw[-stealth] (A)-- node[left] {\small $d$} (B);
  \draw[-stealth] (B)-- node [below] {\small $\varphi$} (D);
  \draw[-stealth] (A)-- node [above] {\small $1$} (AS);
  \draw[-stealth] (AS)-- node [right] {\small $d^{\prime}$} (D);
\end{tikzpicture}
\end{equation*}
where $1$ is the identity mapping on $A$. This means
\begin{equation*}
\forall f, g , h\in A:\qquad \varphi (f\centerdot dg\centerdot h) = f\centerdot d^{\prime}g\centerdot h.
\end{equation*}
\end{definition}
In the example (\ref{e}), two FODCaluli $\Omega_{p}^{1}(A)$ and $\Gamma_{q}^{1}(A)$ are isomorphic iff $p(x) = q(x)$.\\
Also, notice that if in example (\ref{e}) we choose $p(x) = x$, then we recover the usual differential calculus over $A = \mathbb{C}[x]$; e.g;
\begin{equation*}
dx^{2} = x\centerdot dx + dx\centerdot x = x\centerdot dx + x\centerdot dx = 2x\centerdot dx;\cdots\;etc.
\end{equation*}
Next, we would like to define the analogue of the derivative operator $\partial$ as well. Referring to the example (\ref{e}), we can state the following definition.
\begin{definition}
Because $dx$ is a right A-module basis for $\Omega_{p}^{1}(A)$, for every $f\in A$ there exists a unique element, $\partial_{x}f\in A$, such that
\begin{equation}
df = dx\centerdot\partial_{x}f \equiv dx\centerdot\dfrac{\partial f}{\partial x}
\end{equation}
It follows from the uniqueness property due to the freeness on the right in $\Omega_{p}^{1}(A)$, that the correspondence
\begin{equation}
\partial_{x}: \left\lbrace\begin{array}{rl}A&= A \\
f&= \partial_{x}f
\end{array}\right.
\end{equation}
is indeed a well-defined and injective mapping, that is called the derivative operator and $\partial_{x}f$ is called the derivative of $f$.
\end{definition}
We can now verify that
\begin{enumerate}[(1)]
\item $\partial_{x}$ is an $\mathbb{F}$-linear mapping:
\begin{align*}
d(\alpha_{1}f_{1} + \alpha_{2}f_{2}) &= \alpha_{1}df_{1} + \alpha_{2}df_{2}\qquad\qquad \qquad since\;d\;is\;\mathbb{F}-linear,\\
&= \alpha_{1}dx\centerdot\partial_{x}f_{1} + \alpha_{2}dx\centerdot\partial_{x}f_{2}, \\
&= dx\centerdot (\alpha_{1}\partial_{x}f_{1} + \alpha_{2}\partial_{x}f_{2}).
\end{align*}
on the other hand we can also write
\begin{equation*}
d(\alpha_{1}f_{1} + \alpha_{2}f_{2}) := dx\centerdot\partial_{x}(\alpha_{1}f_{1} + \alpha_{2}f_{2}).
\end{equation*}
Using the uniqueness property (i.e; the freeness on the right) we conclude that
\begin{equation*}
\partial_{x}(\alpha_{1}f_{1} + \alpha_{2}f_{2}) = \alpha_{1}\partial_{x}f_{1} + \alpha_{2}\partial_{x}f_{2}.
\end{equation*}
\item $\partial_{x}1 = 0$ follows from $d1 = 0$.
\item $\partial_{x}$ does not obey the simple Leibniz rule. In the example (\ref{e}) for instance we may write 
\begin{align*}
d(f_{1}f_{2}) &= df_{1}\centerdot f_{2} +f_{1}\centerdot df_{2} = dx\centerdot\partial_{x}f_{1}\centerdot f_{2} + f_{1}\centerdot dx\centerdot\partial_{x}f_{2},\\
&= dx\centerdot\partial_{x}f_{1}\centerdot f_{2} + dx\centerdot f_{1}(p(x))\centerdot\partial_{x}f_{2} \\
&= dx\centerdot\left[ \partial_{x}f_{1}\centerdot f_{2} + f_{1}(p(x))\centerdot\partial_{x}f_{2}\right] . 
\end{align*}
On the other hand we may also write
\begin{equation*}
d(f_{1}f_{2}) = dx\centerdot\partial_{x}(f_{1}f_{2}).
\end{equation*}
Using the uniqueness property (i.e. the freeness on the right) we conclude that
\begin{equation}
\partial_{x}(f_{1}f_{2}) = \partial_{x}f_{1}\centerdot f_{2} + f_{1}(p(x))\centerdot\partial_{x}f_{2}. \label{y}
\end{equation}
This is not a simple derivation of $A = \mathbb{C}[x]$, but rather it is the Leibniz rule twisted by an endomorphism $A$, i.e. a twisted derivation, where the twist is effected by the substitution homomorphism of $A$ defined by $p(x)$.
\end{enumerate}
\begin{remark}
\begin{enumerate}[(a)]
\item The mapping 
\begin{equation*}
p:\left\lbrace\begin{array}{rl}\mathbb{C}[x]&\longrightarrow \mathbb{C}[x] \\
f(x)&\longmapsto f(p(x))
\end{array}\right.
\end{equation*}
where $p(x)$ is a chosen polynomial in $\mathbb{C}[x]$, is an algebra homomorphism, called the substitution homomorphism determined by $p(x)$.
\item Let $\alpha \in \mathrm{End}(A) := \mathrm{Hom}_{A}(A, A)$, be an algebra endomorphism. An $\alpha$-derivation of $A$ into $A$ is an $\mathbb{F}$-linear mapping $\delta: A\longrightarrow A$, such that
\begin{equation}
\forall a, b\in A:\qquad \delta (ab) = \delta a\centerdot b+\alpha (a)\centerdot \delta b.
\end{equation}
Because $\alpha (1) = 1$, it follows that $\delta (1) = 0$
\end{enumerate}
\end{remark}
Therefore, $\partial_{x}$ in example (\ref{e}) is a p-derivation of $A$ and not simply a derivation of $A$. From the computations in example (\ref{e}), we can explicitly compute $\partial_{x}$:
\begin{align}
\nonumber dx^{n}&= \sum_{i+j= n-1}x^{i}\centerdot dx\centerdot x^{j} = \sum_{i+j= n-1}dx\centerdot (p(x))^{i}x^{j},\\
\nonumber &=dx\centerdot \left(  \sum_{i+j =n-1}(p(x))^{i}x^{j}\right),\\
&\Longrightarrow \partial_{x}(x^{n})= \sum_{i+j= n-1}(p(x))^{i}x^{j}.
\end{align}
Again this formula reduces to the classical case $\partial_{x}(x^{n}) = nx^{n-1}$, if we choose $p(x) = x$, in which case the endomorphism defined by $p(x)$ becomes the identity endomorphism $1$, i.e. $p(x) = x$.
\begin{exmp}
Let us consider the example (\ref{e}) but take $p(x) = qx$, where $q\in\mathbb{C}, q\neq 1$. Then the $A$-$A$ bimodule $\Omega^{1}(A)$ is given by the equation 
\begin{equation}
f(x)\centerdot dx := dx\centerdot f(qx). \label{t}
\end{equation}
Let us take $f(x) = x+x^{2}$ and compute
\begin{align}
\nonumber df(x) &= d(x+ x^{2}) = dx +dx\centerdot x+ x\centerdot dx = dx + dx\centerdot x+ dx\centerdot q(x),\\
\nonumber &= dx\centerdot (1+x+qx)\\
\nonumber &\equiv dx\centerdot \partial_{x}(x+x^{2}),\\
&\Longrightarrow \partial_{x}(x+ x^{2}) = 1+x +qx. \label{r}
\end{align}
Clearly as $q\longrightarrow 1$, we recover the classical case: $\partial_{x}(x + x^{2}) = 1+ 2x$.\vspace{.5cm}\\
Let us for the moment go back to the classical case, and ask the question: What kind of operation would give a result like the relation (\ref{r}) in the classical case?\\
To answer this question we need the following definition.
\begin{definition}
Let $q\neq 1$. The q-differentiation operator, $D_{q}$, is defined by 
\begin{equation}
D_{q}f(x) = \dfrac{f(x) -f(qx)}{x-qx}.
\end{equation}
Note that as $q\longrightarrow 1$, $D_{q}\longrightarrow\partial_{x}$ of the classical case.
\end{definition}
\begin{result}
$D_{q}f(x)$ in the classical case can be represented by the following infinite series
\begin{equation}
D_{q}f(x) = \sum_{n=0}^{\infty}\dfrac{(q-1)^{n}}{(n+1)!}x^{n}\frac{d^{n+1}}{dx^{n+1}}f(x).
\end{equation}
\end{result}
\begin{verif-non}
Let us write 
\begin{equation*}
f(qx) = f[x+ (qx -x)].
\end{equation*}
and as we are working with the classical case, we can apply the Taylor expansion to write
\begin{align*}
f(qx) &= f[x+ (qx-x)] = f(x) + (qx-x)\frac{df(x)}{dx} + \dfrac{(qx-x)^{2}}{2!}\frac{d^{2}f(x)}{dx^{2}} + \dfrac{(qx-x)^{3}}{3!}\frac{d^{3}f(x)}{dx^{3}} + \cdots \\
&= f(x) + (q-1)x\frac{df}{dx} + \dfrac{(q-1)^{2}}{2!}x^{2}\frac{d^{2}f(x)}{dx^{2}} + \dfrac{(q-1)^{3}}{3!}x^{3}\frac{d^{3}f}{dx^{3}} +\cdots \\
\Longrightarrow \dfrac{f(x) - f(qx)}{x-qx} &= \dfrac{f(qx) - f(x)}{(q-1)x} = \frac{df}{dx} + \dfrac{q-1}{2!}x\frac{d^{2}f}{dx^{2}} + \dfrac{(q-1)^{2}}{3!}x^{2}\frac{d^{3}f}{dx^{3}} + \cdots \\
&= \sum_{n=0}^{\infty}\dfrac{(q-1)^{n}}{(n+1)!}x^{n}\frac{d^{n+1}}{dx^{n+1}}f(x).
\end{align*}
\end{verif-non}
It follows that 
\begin{equation*}
D_{q}(x+x^{2}) = \frac{df}{dx} + \dfrac{q-1}{2!}x\frac{d^{2}f}{dx^{2}} + 0+ \cdots = (1+2x) + \dfrac{q-1}{2}x \centerdot 2 = 1+x+qx.
\end{equation*}
which is the same as $\partial_{x}(x+ x^{2})$ in the case of the relations (\ref{t}) and (\ref{r}).\\
It follows that $\partial_{x}$ as defined for the bimodule defined by the relation (\ref{t}) is just the operator $D_{q}$; that is $\partial_{x}$ is the q-differentiation operator.\\
$D_{q}$ operator satisfies the q-analogue of the Leibniz rule, i.e.
\begin{align}
\nonumber D_{q}[f_{1}(x)f_{2}(x)] &= \dfrac{f_{1}(x)f_{2}(x) - f_{1}(qx)f_{2}(qx)}{x(1-q)},\\
\nonumber &= \dfrac{[f_{1}(x) - f_{1}(qx)]f_{2}(x) + f_{1}(qx)[f_{2}(x) - f_{2}(qx)]}{x(1-q)},\\
\Longrightarrow D_{q}[f_{1}(x)f_{2}(x)] &= D_{q}f_{1}(x)\centerdot f_{2}(x) + f_{1}(qx)D_{q}f_{2}(x).
\end{align}
Analogously, using the relation (\ref{y}), we can write
\begin{equation*}
\partial_{x}(f_{1}(x)f_{2}(x)) = \partial_{x}f_{1}(x)\centerdot f_{2}(x) + f_{1}(qx) \centerdot \partial_{x}f_{2}(x).
\end{equation*}
We see that a non-symmetric bimodule structure on $\Omega_{1}(A)$ results into very interesting and rather surprising differential calculi.
\end{exmp}
\begin{exmp}
Another interesting example is obtained by putting $p(x) = x+c,\; c\neq 0$, in the example (\ref{e}). In this case we obtain
\begin{equation}
f(x)\centerdot dx = dx \centerdot f(x+c).
\end{equation}
As a specific case let us again take $f(x) = x + x^{2}$; then 
\begin{equation*}
df(x) = d(x+ x^{2}) = dx +dx\centerdot x + x\centerdot dx = dx + dx\centerdot x + dx\centerdot (x+c) = dx[(1+c) +2x].
\end{equation*}
Comparing with $df(x) = dx\centerdot \partial_{x}f$ and using the uniqueness property (i.e. freeness of $\Omega^{1}(A)$ on the right), we obtain
\begin{equation*}
\partial_{x}(x+x^{2}) = (1+c) + 2x.
\end{equation*}
Clearly as $c\longrightarrow 0$, we obtain the classical case. Moreover, writing
\begin{equation*}
\partial_{x}(x+x^{2}) = (1+c) +2x = \frac{1}{c}[(x+c) + (x+c)^{2} - x - x^{2}] = \dfrac{f(x+c) -f(x)}{c}
\end{equation*}
where $f(x) = x+x^{2}$ helps us to recognize that $\partial_{x}$ is the following classical operator
\begin{equation}
\partial_{x}f(x) = \dfrac{f(x+c) -f(x)}{c}
\end{equation}
In the limit $c\longrightarrow 0$ we recover the usual differentiation of the classical case.
\end{exmp}
\section{First order differential calculi on associative algebra, based on inner derivation}
\begin{definition}
Let $A$ be an associative algebra. 
\begin{enumerate}[(1)]
\item For any element $f\in A$ the mapping
\begin{equation}
ad_{f} := d_{f} : \left\lbrace\begin{array}{rl}A&\longrightarrow A\\
g&\longmapsto [g, f] := gf-fg
\end{array}\right.
\end{equation}
is a derivation of $A$ into $A$ for,
\begin{equation*}
\forall g, h\in A: ad_{f}(gh) = [gh, f] = g\centerdot [h, f] +[g,f]\centerdot h =ad_{f}(g)\centerdot h+ g\centerdot ad_{f}(h),
\end{equation*}
which shows that $ad_{f}$ satisfies the Leibniz rule. It is easily seen to be linear for,
\begin{equation*}
ad_{f}(\alpha g+\beta h) = [\alpha g+\beta h, f] = \alpha [g, f] +\beta [h, f] = \alpha \;ad_{f}(g) +\beta \;ad_{f}(h).
\end{equation*}
Hence $ad_{f}: A\longrightarrow A$ is a derivation of $A$ into $A$.
\item Let $M$ be an $A$-$A$ bimodule and $m\in M$. The mapping
\begin{equation}
ad_{m} := d_{m}: \left\lbrace\begin{array}{rl}A&\longrightarrow M \\
f &\longmapsto [f, m] := f\centerdot m - m\centerdot f
\end{array}\right.
\end{equation}
satisfies the Leibniz rule
\begin{align*}
d_{m}(fg) &= [fg, m] = (fg)\centerdot m - m\centerdot (fg) = f\centerdot (g\centerdot m) - (m\centerdot f)\centerdot g,\\
&=f \centerdot (g\centerdot m)  \underbrace{-f\centerdot (m\centerdot g) + f\centerdot (m\centerdot g)}_{=0} - (m\centerdot f)\centerdot g,\\
&= \underbrace{f\centerdot (g\centerdot m) - f\centerdot (m\centerdot g)}_{f\centerdot ad_{m}(g)} +\underbrace{f\centerdot (m\centerdot g) - 
(m\centerdot f)\centerdot g}_{ad_{m}(f)\centerdot g}.
\end{align*}
which is the Leibniz rule. Clearly $ad_{m}$ is linear as can be easily verified. We conclude that 
\begin{equation*}
d_{m} := ad_{m} = [\centerdot , m].
\end{equation*}
is a derivation of $A$ into $M$. It is called an inner derivation of $A$ into $M$.
\end{enumerate}
\end{definition}
\begin{exmp}\label{u}
Let us consider the algebra $A = \mathbb{C}[x]$ again. Choose an element $f\in A$ and let $d_{f}: A\longrightarrow A$ be an inner derivation of $A$ into $A$; $d_{f} = [\centerdot , f]$. Let us denote $Im(d_{f})$ in $A$ by $d_{f}(A)$, and consider the $A$-$A$ bimodule
\begin{equation}
\Omega_{p, f}^{1}(A) := A\centerdot d_{f}(A)\centerdot A,\quad p\in A.
\end{equation}
such that this $A$-$A$ bimodule is free from the right as an A-module, and the bimodule structure is specified by $p(x)\in A$ according to the following relation
\begin{equation}
\forall g, h\in A:\qquad g\centerdot d_{f}(h) = d_{f}(h)\centerdot g(p(x)).
\end{equation}
We can now verify that 
\begin{align*}
\forall h, g\in A: \qquad d_{f}(gh)&= (d_{f}g)\centerdot h + g\centerdot (d_{f}h) = (d_{f}g)\centerdot h + (d_{f}h)\centerdot g(p(x)),\\
&= (d_{f}x\centerdot \partial_{x}(g))\centerdot h + (d_{f}x\centerdot \partial_{x}(h))\centerdot g(p(x)),
\end{align*}
where $\partial_{x}(g)$ and $\partial_{x}(h)$ are unique elements of $A$,
\begin{equation*}
d_{f}(gh) = d_{f}x\centerdot (\partial_{x}(g)\centerdot h + \partial_{x}(h)\centerdot g(p(x))).
\end{equation*}
On the other hand we can also write 
\begin{equation*}
d_{f}(gh) = d_{f}x\centerdot\partial_{x}(gh)
\end{equation*}
where $\partial_{x}(gh)$ is a unique element in $A$, by the freeness on the right. Equating the right hand sides of these two expressions, we obtain
\begin{equation}
\partial_{x}(gh) = \partial_{x}(g)\centerdot h + \partial_{x}(h)\centerdot g(p(x)).
\end{equation}
The bimodule $\Omega_{p, f}^{1}(A) := A\centerdot d_{f}(A)\centerdot A$ togather with the derivation $d_{f} := [\centerdot , f]$ as the differential mapping is a FODC over $A$.
\end{exmp}
\begin{exmp}
Consider the FODC over $A = \mathbb{C}[x]$, $\Omega_{p}^{1}(A)$, as was constructed in example (\ref{e}). Consider this A-A bimodule and choose an element $m\in \Omega_{p}^{1}(A),\; m\neq 0$ which we call a one form, and consider the mapping
\begin{equation}
d_{m} : \left\lbrace\begin{array}{rl}A&\longrightarrow \Omega_{p}^{1}(A) \\
f &\longmapsto [f, m] := f\centerdot m - m\centerdot f
\end{array}\right.
\end{equation}
Clearly $d_{m}$ is a linear mapping. Moreover, as in the example (\ref{u}) we can verify that $d_{m}$ satisfies the Leibniz rule. It is, therefore, a derivation of $A$ into the bimodule $\Omega_{p}^{1}(A)$. Denoting the image of $d_{m}$ in $\Omega_{p}^{1}(A)$ by $d_{m}(A)$, we consider the A-A bimodule
\begin{equation}
\Gamma_{p, m}^{1}(A) = A\centerdot d_{m}(A)\centerdot A.
\end{equation} 
which is free on the right, togather with the mapping $d_{m}$ is a FODC over $A$ with the differential mapping $d_{m}$.\\
Let us choose $m= gdh$, a one form in $\Omega_{p}^{1}(A)$, then we may compute for every $f\in A$:
\begin{align}
\nonumber d_{m}f := [f, m] &= [f, gdh] = f\centerdot gdh -gdh\centerdot f = f\centerdot g\centerdot dx\centerdot\partial_{x}h - g\centerdot dx\centerdot (\partial_{x}h)\centerdot f,\\
\nonumber &= g\centerdot dx\centerdot (\partial_{x}h)\centerdot f(p(x)) - g\centerdot dx\centerdot (\partial_{x}h)\centerdot f\\
\nonumber &= dx\centerdot g(p(x))\centerdot (\partial_{x}h)\centerdot f(p(x)) - dx\centerdot g(p(x))\centerdot (\partial_{x}h)\centerdot f\\
&= dx\centerdot g(p(x))\centerdot (\partial_{x}h)\centerdot [f(p(x)) - f(x)]. \label{i}
\end{align}
On the other hand $d_{m}f\in \Omega_{p}^{1}(A)$ can be uniquely written as $d_{m}f = dx\centerdot (\partial_{x}f)$, where $\partial_{x}f$ is a unique element of $A$. Comparing this with the relation (\ref{i}), by freeness of $\Omega_{p}^{1}(A)$ on the right (and hence that of $\Gamma_{p, m}^{1}(A)$ on the right), we obtain the partial derivative
\begin{equation}
\partial_{x}f = g(p(x))\centerdot (\partial_{x}h)\centerdot [f(p(x)) - f(x)].
\end{equation}
\end{exmp}
\section{First order coordinate differential calculi (FOCDC) over associative algebra}
Now it is important to work with generators and relations for the associative algebra involved. We call an associative algebra $A$ given by 
\begin{equation}
A= \bigslant{\mathbb{F}\langle x^{1}, \ldots , x^{n}\rangle}{(f_{1}, \ldots , f_{p})} = \bigslant{\mathcal{F}}{I}.
\end{equation}
a coordinate algebra, because we would like to interpret $A$ as the algebra of polynomial functions $A = Func(X)$ over some noncommutative space $X$, which we usually call a quantum space. This resembles the situation in algebraic geometery. Homogeneous ideals corresponding to graded algebras (projective case). Roughly speaking coordinate algebras are quantum analogue of algebraic varieties.\\
We will see in this section that a noncommutative differential calculus is best handled by means of commutation relations among the generators of the algebra  and their differentials. That is why it is crucial to work with presentations of the algebras of interest.\vspace{.5cm}\\
Let  $\mathcal{F} = \mathbb{F}\langle x^{1}, \ldots , x^{n}\rangle$ be the free algebra on the generators $x^{1}, \ldots , x^{n}$, over a field $\mathbb{F}$ and let $I\vartriangleleft \mathcal{F}$ be a two sided ideal in $\mathcal{F}$. The quotient algebra $A = \bigslant{\mathcal{F}}{I}$ is of basic interest here:
\begin{align}
\nonumber A &= \bigslant{\mathcal{F}}{I} = \bigslant{\mathbb{F}\langle x^{1}, \ldots ,x^{n}\rangle}{\left( f_{1}(x^{1}, \ldots , x^{n}), \ldots , f_{p}(x^{1}, \ldots , x^{n})\right) } ,\\
&= \mathbb{F}\langle \bar{x}^{1}, \bar{x}^{2}, \ldots , \bar{x}^{n}\vert f_{i}(\bar{x}^{1}, \ldots , \bar{x}^{n}) = 0,\quad i= 1, \ldots , p \rangle \label{o}
\end{align}
The canonical algebra epimorphism $\eta_{_{I}}$ is 
\begin{equation*}
\eta_{_{I}}:  \left\lbrace\begin{array}{rl}\mathcal{F}&\longrightarrow \bigslant{\mathcal{F}}{I},\qquad Ker(\eta_{_{I}}) = I,\\
\bar{x_{k}} &:=  x_{k} + I = \eta_{_{I}}(x_{k}),\qquad k= 1, \ldots , p.
\end{array}\right.
\end{equation*}
when there is no danger of confusion, we shall simply write the relation (\ref{o}) as
\begin{equation}
A = \bigslant{\mathcal{F}}{I} = \mathbb{F}\left\langle x^{1}, \ldots , x^{n}\vert f_{i}(x^{1}, \ldots , x^{n}) = 0,\; i= 1, \ldots , p\right\rangle .
\end{equation}
Let us write the bimodule of one form as $\Omega_{d}^{1}(A) = A\centerdot d(A)\centerdot A$ with the assumption that this bimodule is a free right A-module. Then the pair $\left( \Omega_{d}^{1}(A), d\right) $ is called a first order coordinate differential calculu (FOCDC, for short) over $A$. The differential map $d$ is called a coordinate differential or a free differential.\\
The freeness of $\Omega_{d}^{1}(A)$ on the right allows us to write 
\begin{equation}
\forall f;\qquad df= dx^{i}\centerdot f_{i}.
\end{equation}
in a unique manner, for unique elements $f_{i}\in A,\; i= 1, \ldots , n$. For this reson $d$ is called a "free" differential. This uniqueness property allows us to define the partial derivatives (= vector fields), linear mapping $\partial_{k}: A\longrightarrow A$ by the formula
\begin{equation}
\forall f\in A:\qquad df = dx^{k}\centerdot\partial_{k}f. \label{p}
\end{equation} 
where summation over $k$ from $1$ to $n$ is assumed and where $\partial_{k}f,\; k= 1, \ldots ,n$ are unique elements of $A$. It follows immediately from the relation (\ref{p}) that 
\begin{equation}
dx^{i} = dx^{k}\centerdot\partial_{k}x^{i}\Longrightarrow \partial_{k}x^{i} = \delta_{k}^{i}.\label{f}
\end{equation}
At this stage we can move along two different paths.
\begin{enumerate}[\underline{Path}~1.]
\item As above use the differential map and construct the FOCDC, $\left( \Omega_{d}^{1}(A), d\right) $, In this approach the partial derivatives are computed using the freeness of the bimodule $\Omega_{d}^{1}(A)$ on the right, as we did above. This is called a derivation-based calculus ( or differential based calculus ).
\item This approach is based on derivation $\partial $ instead of $d$. That is the differential map $d$ will be defined as a consequence of the properties of $\partial$. This approach is called a derivative based FOCDC, as it is usual in the classical case. We shall see that this method is not as general as the first one when we try to connect to geometery.
\end{enumerate}
\begin{enumerate}[\underline{\textbf{Method}}~1.]
\item \subsubsection*{Differential based FOCDC over A}
The module $\Omega_{d}^{1}(A)$ of one form is completely specified if we define the left action of $A$ on it. Let us take $dx^{i}\in\Omega_{d}^{1}(A)$ and multiply it on the left by an element $f\in A$. Because $f\centerdot dx^{i}\in\Omega_{d}^{1}(A)$, by the freeness of $\Omega_{d}^{1}(A)$ on the right, we conclude that there exist a unique set of elements $T_{k}^{i}(f)\in A,\; i, k = 1, \ldots , n$, such that
\begin{equation}
\forall i=1, \ldots , n,\; \forall f\in A:\qquad f\centerdot dx^{i} = dx^{k}\centerdot T_{k}^{i}(f). \label{a}
\end{equation}
where the summation over repeated up and down indices is assumed. This relation completely determines the module $\Omega_{d}^{1}(A)$ of one form if we know the elements $T_{k}^{i}(f)\in A,\; i, k= 1, \ldots , n$. \\
Using the relation (\ref{a}) we can now write,
\begin{equation*}
\forall f, g\in A:\quad (fg)\centerdot dx^{i} = f\centerdot (g\centerdot dx^{i}) = f\centerdot dx^{j}\centerdot T_{j}^{i}(g) = dx^{k}\centerdot T_{k}^{j}(f)T_{j}^{i}(g).
\end{equation*}On the other hand, using the relation (\ref{p}), we can directly write 
\begin{equation*}
(fg)\centerdot dx^{i} = dx^{k}\centerdot T_{k}^{i}(fg).
\end{equation*}
Again by the freeness of $\Omega_{d}^{1}(A)$ on the right, we conclude 
\begin{equation}
T_{k}^{i}(fg) = T_{k}^{j}(f)T_{j}^{i}(g).
\end{equation}
which shows that the mapping 
\begin{equation}
T: \left\lbrace\begin{array}{rl}A&\longrightarrow M_{n}(A)\\
f &\longmapsto T(f)
\end{array}\right. \label{s}
\end{equation}
is an algebra homomorphism.\\
This argument also demonstrates that, conversely, given any algebra homomorphism $T$ by the relation (\ref{s}), one obtains a unique bimodule $\Omega_{d}^{1}(A)$ of one forms, and hence a unique FOCDC $\left( \Omega_{d}^{1}(A), d\right) $ over $A$.\\
Hence there exists a one to one correspondence
\begin{equation}
Bimodule\;\; \Omega_{d}^{1}(A)\overset{1:1}{\longleftrightarrow} \mathrm{Hom}_{Alg}(A, M_{n}(A)). \label{c}
\end{equation}
Applying the relation (\ref{a}) to the generators of $A$ (which are to be thought of as the coordinates of a noncommutative space) we obtain
\begin{equation}
x^{j}\centerdot dx^{i} = dx^{k}\centerdot T^{i}_{k}(x^{j}). \label{b}
\end{equation}
$T= (T^{i}_{k})$ is a $n\times n$ matrix belonging to $M_{n}(A)$, in which $i$ is the column and $k$ is the row index.\\
We can now go on to compute the derivatives $\partial_{i},\; i= 1, \ldots ,n$. As we shall see the derivatives do not satisfy the simple Leibniz rule but a twisted one.
\begin{align*}
\forall f, g\in A:\quad d(fg) &= df\centerdot g+ f\centerdot dg = (dx^{i}\centerdot\partial_{i}f)\centerdot g + f\centerdot (dx^{i}\centerdot\partial_{i}g),\\
&= (dx^{i}\centerdot\partial_{i}f)\centerdot g + (f\centerdot dx^{i})\centerdot\partial_{i}g,\\
&= (dx^{k}\centerdot\partial_{k}f)\centerdot g+ (dx^{k}\centerdot T^{i}_{k}(f))\centerdot\partial_{i}g,\\
&= dx^{k}\centerdot \left( \partial_{k}f\centerdot g+ T^{i}_{k}(f)\centerdot\partial_{i}g\right). 
\end{align*}
But also we can write
\begin{equation*}
d(fg) = dx^{k}\centerdot\partial_{k}(fg).
\end{equation*}
Equating these two expressions and using the freeness of $\Omega_{d}^{1}(A)$ on the right, we conclude
\begin{equation}
\partial_{k}(fg) = \partial_{k}f\centerdot g + T^{i}_{k}(f)\centerdot\partial_{i}g.\label{d}
\end{equation}
This shows that 
\begin{equation*}
\partial_{k}: A\longrightarrow A,\qquad k= 1, \ldots ,n.
\end{equation*}
is a derivation of $A$ into $A$ twisted by the homomorphism $T$. There is however one property that $\partial_{k}$ should possess for it to be a twisted homomorphism, and that is
\begin{equation*}
\partial_{k} =0,\qquad k= 1, \ldots ,n.
\end{equation*}
We now show that this is a consequence of the relation (\ref{d}):\\
Using this relation we can write
\begin{equation*}
\partial_{k}(1) = \partial_{k}(1\centerdot 1) = (\partial_{k}1)\centerdot 1 + T^{i}_{k}(1)\centerdot \partial_{i}(1),
\end{equation*}
But $T^{i}_{k}(1) = \delta^{i}_{k}$, so this yields
\begin{equation*}
\partial_{k}(1) = \partial_{k}(1) + \delta^{i}_{k}\centerdot\partial_{i}(1) = \partial_{k}(1) + \partial_{k}(1)\Longrightarrow \partial_{k}(1) = 0.
\end{equation*}
As a final point, using the property of the free differential (or the coordinate differential), we conclude that $\partial_{k}(x^{i}) = \delta^{i}_{k}$, which was obtained in the relation (\ref{f}).
\item \subsubsection*{Derivative approach}
In this method we shall use the partial derivatives to define a FOCDC over $A$. For this purpose we must go through the following steps.
\begin{enumerate}[(a)]
\item Let $\partial_{k}: A\longrightarrow A,\; k= 1, \ldots , n$, be linear mappings which satisfy
\begin{align}
\partial_{k}(x^{i}) &= \delta^{i}_{k},\quad \forall i, k= 1, \ldots , n. \label{j}\\
\partial_{k}(fg) &= \partial_{k}(f)\centerdot g+ T^{i}_{k}(f)\centerdot \partial_{i}(g). \label{g}
\end{align}
where for each $f\in A,\; T^{i}_{k}(f)$ are some elements of $A$. We shall use the relation (\ref{g}) and the associativity of product in $A$ to show that $T$ is an algebra homomorphism:
\begin{equation*}
\forall f, g, h\in A:\quad\partial_{k}(f(gh)) = \partial_{k}((fg)h).
\end{equation*}
on using the relation (\ref{g}) we can write
\begin{equation*}
\partial_{k}f\centerdot (gh) + T^{i}_{k}(f)\centerdot\partial_{i}(gh) = \partial_{k}(fg)\centerdot h + T^{i}_{k}(fg)\centerdot\partial_{i}h.
\end{equation*}
Applying the relation (\ref{g}) again we can write
\begin{align*}
&\partial_{k}f\centerdot (gh) + T^{i}_{k}(f)\centerdot [\partial_{i}(g)\centerdot h + T^{j}_{i}(g)\centerdot\partial_{j}h] = [\partial_{k}(f)\centerdot g + T^{i}_{k}(f)\centerdot\partial_{i}g]\centerdot h + T^{i}_{k}(fg)\centerdot\partial_{i}h.\\
&\Longrightarrow\partial_{k}f\centerdot(gh) +T^{i}_{k}(f)\centerdot(\partial_{i}g)\centerdot h + T^{i}_{k}(f)T^{j}_{i}(g)\centerdot\partial_{j}h = \\
& = \partial_{k}f\centerdot(gh) +T^{i}_{k}(f)\centerdot(\partial_{i}g)\centerdot h +T^{j}_{k}(fg)\centerdot\partial_{j}h\\
&\Longrightarrow T^{i}_{k}(f)T^{j}_{i}(g)\centerdot\partial_{j}h = T^{j}_{k}(fg)\centerdot\partial_{j}h.
\end{align*}
This should hold for all $f, g, h$ and in particular for $h = x^{l}$. Using this choice we obtain
\begin{equation*}
T^{i}_{k}(f)T^{j}_{i}(g)\centerdot\partial_{j}x^{l} = T^{j}_{k}(fg)\centerdot\partial_{j}x^{l}.
\end{equation*}
Now, by the relation (\ref{j}) $\partial_{j}x^{l} = \delta^{l}_{j}$, and hence we obtain
\begin{equation*}
T^{l}_{k}(fg) = T^{i}_{k}(f)T^{l}_{i}(g).
\end{equation*}
and this proves that the mapping
\begin{equation}
T: A\longrightarrow M_{2}(A),\quad T= (T^{i}_{j}),\quad i= column\;index,\quad j= row\;index.
\end{equation}
is an algebra homomorphism.
\item Consider the set of $n$ symbols $\lbrace d^{\prime}x^{1}, d^{\prime}x^{2}, \ldots , d^{\prime}x^{n}\rbrace$ and let $\Gamma_{d^{\prime}}^{1}(A)$ be the free right A-modul on this set, that is 
\begin{equation}
\Gamma_{d^{\prime}}^{1}(A) = \sum_{i=1}^{n}d^{\prime}x^{i}\centerdot A\cong \underbrace{A\oplus A\oplus\ldots\oplus A}_{n-fold}. \label{l}
\end{equation}
Let us pick $d^{\prime}x^{i}\in \Gamma_{d^{\prime}}^{1}(A)$ and multiply it on the left by $f\in A$. The left multiplication by $f\in A$ acts in $\Gamma_{d^{\prime}}^{1}(A)$ by an endomorphism of this module. We know that the endomorphism algebra of the relation (\ref{l}) is $M_{n}(A)$. We conclude that 
\begin{equation}
f\centerdot d^{\prime}x^{i} = d^{\prime}x^{k}\centerdot T^{i}_{k}(f).\label{z}
\end{equation}
where we have deliberately chosen the endomorphism $T: A\longrightarrow M_{n}(A)$ of the method 1. the relation (\ref{z}) defines the left action of $A$ in $\Gamma_{d^{\prime}}^{1}(A)$, and hence fixes the bimodule structure.\\
let us next consider the following mapping
\begin{equation}
d^{\prime}: \left\lbrace\begin{array}{rl}A&\longrightarrow \Gamma_{d^{\prime}}^{1}(A)\\
f &\longmapsto d^{\prime}x^{k}\centerdot\partial_{k}f \label{x}
\end{array}\right.
\end{equation}
where $\partial_{k}: A\longrightarrow A$ is defined by the relations (\ref{j}) and (\ref{g}). This mapping is $\mathbb{F}$-linear for
\begin{align*}
\forall \alpha_{1}, \alpha_{2}\in\mathbb{F},\; \forall f_{1}, f_{2}\in A&:\\
 d^{\prime}(\alpha_{1}f_{1} + \alpha_{2}f_{2}) &= d^{\prime}x^{k}\centerdot\partial_{k}(\alpha_{1}f_{1} + \alpha_{2}f_{2}),\\
 &= \alpha_{1}d^{\prime}x^{k}\centerdot\partial_{k}(f_{1}) + \alpha_{2}d^{\prime}x^{k}\centerdot\partial_{k}(f_{2}) = \alpha_{1}d^{\prime}f_{1} + \alpha_{2}d^{\prime}f_{2}.
\end{align*}
We now verify that $d^{\prime}$ as given by the relation (\ref{x}) satisfies the Leibniz rule:
\begin{align*}
d^{\prime}(fg) &= d^{\prime}x^{k}\centerdot\partial_{k}(fg) = d^{\prime}x^{k}\centerdot\left[ \partial_{k}(f)\centerdot g + T^{j}_{k}(f)\centerdot\partial_{j}g\right],\\
&= \left( d^{\prime}x^{k}\centerdot\partial_{k}f\right)\centerdot g + \left( d^{\prime}x^{k}\centerdot T^{j}_{k}(f)\right)\centerdot\partial_{j}g,\\
&= d^{\prime}f\centerdot g + (f\centerdot d^{\prime}x^{j})\centerdot\partial_{j}g = d^{\prime}f\centerdot g + f\centerdot (d^{\prime}x^{j}\centerdot\partial_{j}g),\\
&= d^{\prime}f\centerdot g + f\centerdot d^{\prime}g.   
\end{align*}
Hence $d^{\prime}$ is a (coordinate) differential mapping or a free differential.\\
We conclude that the pair $\left( \Gamma_{d^{\prime}}^{1}(A), d^{\prime}\right) $ is a FOCDC over $A$. Moreover since we have used the same $T\in \mathrm{Hom}_{alg}(A, M_{n}(A))$ that was used to construct the FOCDC $(\Omega_{d}^{1}(A), d)$, by what we said in the relation (\ref{c}), these two FOCDC over $A$ are essentially the same. To put it in a more formal language, these two differential calculi are isomorphic. To verify this let us define a mapping of free right A-modules
\begin{equation*}
\varphi :\left\lbrace\begin{array}{rl}\Omega_{d}^{1}(A)&\longrightarrow \Gamma_{d^{\prime}}^{1}(A)\\
dx^{i} &\longmapsto d^{\prime}x^{i},\quad i= 1, \ldots , n
\end{array}\right.
\end{equation*}
since $\varphi$ is given on the generators as indicated. it can be extended to an isomorphism of these free right A-bimodules. It follows that
\begin{align*}
\forall f\centerdot dx^{i}\in \Omega_{d}^{1}(A):\quad \varphi(f\centerdot dx^{i}) &= \varphi(dx^{k}\centerdot T^{i}_{k}(f)) = \varphi(dx^{k})\centerdot T^{i}_{k}(f),\\
&=d^{\prime}x^{k}\centerdot T^{i}_{k}(f) = f\centerdot d^{\prime}x^{i}.
\end{align*}
So, $\varphi$ satisfies the requirement $\varphi (f\centerdot dg\centerdot h) = f\centerdot d^{\prime}g\centerdot h$ for every $f, g, h\in A$, and makes the diagram
\begin{equation*}
\begin{tikzpicture}
  \node (A) {$A$};
  \node (B) [below=of A] {$\Omega_{d}^{1}(A)$};
  \node (AS) [right=of A] {$A$};
  \node (D) [right=of B] {$\Gamma_{d^{\prime}}^{1}(A)$};
  \draw[-stealth] (A)-- node[left] {\small $d$} (B);
  \draw[-stealth] (B)-- node [below] {\small $\varphi$} (D);
  \draw[-stealth] (A)-- node [above] {\small $1$} (AS);
  \draw[-stealth] (AS)-- node [right] {\small $d^{\prime}$} (D);
\end{tikzpicture}
\end{equation*}
commutative. It follows that $\left( \Omega_{d}^{1}(A), d\right) $ and $\left( \Gamma_{d^{\prime}}^{1}(A), d^{\prime}\right) $ are isomorphic.
\end{enumerate}
\end{enumerate}
\subsection{The bimodule of vector fields}
Suppose we have constructed a FOCDC over $A$, say $\left( \Omega_{d}^{1}(A), d\right) $. Let us denote $\Omega_{d}^{1}(A)$ by $M$, for the simpicity of the notations. Recall that $M$ is a free right A-module of one form.\\
Let $M^{\ast}$ be a free left module over $A$, freely generated by the partial derivatives $\partial_{i},\; i=1, \ldots, n$. We may define a right A-module structure on $M^{\ast}$ by the transpose commutation rules 
\begin{equation}
\forall f\in A,\; Y\in M^{\ast};\quad Y\centerdot f \equiv\left( Y^{i}\centerdot\partial_{i}\right)\centerdot f = Y^{i}\centerdot\left( T^{k}_{i}(f)\right)\partial_{k}.  \label{v}
\end{equation}
In this manner we define a bimodule $M^{\ast}$ of vector fields as a dual to a bimodule $M$ of differential forms, togather with a pairing
\begin{equation}
\forall Y\in M^{\ast},\;\forall w= dx^{i}\centerdot w_{i}\in M:\quad \langle Y, w\rangle = Y^{i}w_{i}\in A.
\end{equation}
where we have used 
\begin{equation*}
Y= Y^{i}\centerdot\partial_{i},\; w= dx^{i}\centerdot w_{i}\quad and\quad \langle\partial_{i}, dx^{j}\rangle = \partial_{i}x^{j} = \delta^{j}_{i}.
\end{equation*}
It can be verified that 
\begin{equation}
\langle Y\centerdot f, w\rangle = \langle Y, f\centerdot w\rangle .
\end{equation}
A vector field $Y\in M^{\ast}$ can be characterized as a linear map $Y: A\longrightarrow A$, which satisfies the twisted Leibniz rule
\begin{equation}
\forall f, g\in A,\; \forall Y\in M^{\ast}:\quad Y(fg) = Y(f)\centerdot g + (Y\centerdot f)(g).
\end{equation}
where $Y(f) = Y^{i}\partial_{i}(f)$ and $Y\centerdot f$ is given by the relation (\ref{v}). This is a generalization of the relation (\ref{g}) to the case of an arbitrary vector field $Y$.\vspace{.5cm}\\
Both definition of differential one form and the vector fields essentially depend on the generating vector space $V:=linspan\lbrace x^{1},\ldots , x^{n}\rbrace\subset A$ in the following sense:\\
Let $Z^{k} = p^{k}_{i}x^{i}$ be another basis in the vector space $V$ over $\mathbb{F}$, with the matrix $(p^{k}_{i})\in GL(n, \mathbb{F})$. Then the basis of a bimodule $M$ of differentials $dx^{i}$ and the basis vector field $\partial_{i}\equiv\frac{\partial}{\partial x^{i}}$, undergo the corresponding covariant and contravariant transformation laws, i.e. 
\begin{equation}
dZ^{k} = p^{k}_{i}dx^{i},\quad \frac{\partial}{\partial z^{k}} = q^{i}_{k}\frac{\partial}{\partial x^{i}}.
\end{equation}
where
\begin{equation*}
q^{k}_{l}p^{l}_{i} = \delta^{k}_{i}.
\end{equation*}
which in matrix form is $\left( q^{i}_{k}\right) = \left( (p^{i}_{k})^{t}\right)^{-1}$.
\section{Construction of FOCD Calculi over a given algebra}
We have seen so far that specifing an algebra homomorphism $T: A\longrightarrow M_{n}(A)$, uniquely determines a FOCDC over $A$, by the relation (\ref{b}). So, we are now faced with the problem of how to specify such a homomorphism. Clearly if $A$ is a free algebra, any assignment
\begin{equation}
T: \left\lbrace\begin{array}{rl}\mathcal{F} = \mathbb{F}\langle x^{1}, \ldots , x^{n}\rangle &\longrightarrow M_{n}(\mathcal{F})\\
x^{i} &\longmapsto T(x^{i})\in M_{n}(\mathcal{F})
\end{array}\right.
\end{equation} 
uniquely extends to an algebra homomorphism, and so in this case it is enough to arbitrary choose $n\times n$ matrices $T(x^{i}),\; , i=1, \ldots , n$. \\
However, if the algebra $A$ is not free, then there will be constraints among $T(x^{i}),\; i= 1, \ldots , n$, coming from relations in $A$. To clarify this point, let $I\vartriangleleft\mathcal{F}= \mathbb{F}\langle x^{1},\ldots , x^{n}\rangle$ be a 2-sided ideal in $\mathcal{F}$ say
\begin{equation}
I= \left( R_{l}(x^{1},\ldots , x^{n})\vert l=1, \ldots ,\lambda\right). 
\end{equation}
and let
\begin{equation}
A=\bigslant{\mathcal{F}}{I} = \bigslant{\mathbb{F}\langle x^{1},\ldots ,x^{n}\rangle}{\left( R_{l}\vert l=1,\ldots ,\lambda\right) }.
\end{equation}
where each $R_{l}(x^{1},\ldots ,x^{n})$ is a polynomial in the generators $x^{1}, \ldots , x^{n}$. It is clear that any algebra homomorphism
\begin{equation*}
T: A= \bigslant{\mathcal{F}}{I}\longrightarrow M_{n}(A) = M_{n}\left( \bigslant{\mathcal{F}}{I}\right)\cong\bigslant{M_{n}(\mathcal{F})}{M_{n}(I)}. 
\end{equation*}
must satisfy
\begin{equation}
T(I)\subseteq M_{n}(I). \label{m}
\end{equation}
which can be more explicity written as
\begin{equation}
\forall i, j= 1, \ldots ,n:\quad T^{i}_{j}(I)\subseteq I. \label{n}
\end{equation}
where here $T^{i}_{j}$ are the components of the $n\times n$ matix $M_{n}(A)$. $i$ is the column and $j$ the row indices. We can equivalently write the relations (\ref{m}) and (\ref{n}), as
\begin{equation}
T^{i}_{j}\left( R_{l}(x^{1}, \ldots ,x^{n})\right)\in I\qquad \forall i,j =1, \ldots ,n;\; \forall l=1, \ldots \lambda . \label{qw}
\end{equation}
What has here happened is mentioned in the following:\\
If we choose $T(x^{i})\; i= 1,\ldots ,n$, arbitrarily, we obtain a homomorphism $T:\mathcal{F}\longrightarrow M_{n}(\mathcal{F})$. But $T$ induces $\tilde{T}: A = \bigslant{\mathcal{F}}{I}\longrightarrow M_{n}\left( \bigslant{\mathcal{F}}{I}\right) $ according to the following commutative diagram
\begin{equation}
\begin{tikzpicture}
  \node (A) {$\mathcal{F}$};
  \node (B) [below=of A] {$A= \bigslant{\mathcal{F}}{I}$};
  \node (AS) [right=of A] {$M_{n}(\mathcal{F})$};
  \node (D) [right=of B] {$M_{n}\left( \bigslant{\mathcal{F}}{I}\right)\cong\frac{M_{n}(\mathcal{F})}{M_{n}(I)} $};
  \draw[-stealth] (A)-- node[left] {\small $\eta_{_{I}}$} (B);
  \draw[-stealth] (B)-- node [below] {\small $\tilde{T}$} (D);
  \draw[-stealth] (A)-- node [above] {\small $T$} (AS);
  \draw[-stealth] (AS)-- node [right] {\small $M_{n}(\eta_{_{I}})$} (D);
\end{tikzpicture} \label{qe}
\end{equation}
where $\tilde{T}$ is the homomorphism induced from $T$ in passing to the quotient $A= \bigslant{\mathcal{F}}{I}$, and where $\eta_{_{I}}$ and $M_{n}(\eta_{_{I}})$ are the canonical quotient maps. The commutativity of the diagram requires that the composite mappings
\begin{equation*}
\begin{tikzpicture}
  \node (A) {$I$};
  \node (B) [below=of A] {$\bar{0}$};
  \node (D) [right=of B] {$(0)$};
  \draw[-stealth] (A)-- node[left] {\small $\eta_{_{I}}$} (B);
  \draw[-stealth] (B)-- node [below] {\small $\tilde{T}$} (D);
\end{tikzpicture}
\qquad\qquad and \qquad\qquad
\begin{tikzpicture}
  \node (A) {$I$};
  \node (AS) [right=of A] {$T(I)$};
  \node (D) [right=of B] {$\bigslant{T(I)}{M_{n}(I)}$};
  \draw[-stealth] (A)-- node [above] {\small $T$} (AS);
  \draw[-stealth] (AS)-- node [right] {\small $M_{n}(\eta)$} (D);
\end{tikzpicture}
\end{equation*}
be equal (notice that $T(I)$ is an ideal in $M_{n}(\mathcal{F})$).\\
This implies
\begin{equation}
\bigslant{T(I)}{M_{n}(I)} = (\bar{0})\Longrightarrow T(I)\subseteq M_{n}(I).
\end{equation}
which in component form is the relations (\ref{n}) and (\ref{qw}). In what follows for the sake of simplicity of notations, we shall denote $\tilde{T}$ by $T$and $\bar{x}^{i}$ by $x^{i}$ whenever it causes no confusion.
\begin{definition}
Let $T: \mathcal{F}= \mathbb{F}\langle x^{1}, \ldots , x^{n}\rangle\longrightarrow M_{n}(\mathcal{F})$ be an algebra homomorphism.
\begin{enumerate}[(1)]
\item An ideal $I\vartriangleleft\mathcal{F}$ is called a T-Consistent ideal if it satisfies the relation (\ref{m}) (or its alternatives relations (\ref{n}) and (\ref{qw})).
\item An ideal $I\vartriangleleft\mathcal{F}$ is called T-derivative Consistent if the following conditions hold
\begin{equation}
\forall i=1, \ldots ,n:\quad\partial_{i}(I)\subseteq I.
\end{equation}
where $\partial_{i}$ are the partial derivatives defined by the differential $d$ corresponding to $T$.
\item An ideal $I\vartriangleleft\mathcal{F}$ is said to be \texttt{supported by $T$} if the quotient algebra $A=\bigslant{\mathcal{F}}{I}$ has a FOCDC given by the commutation rules relation (\ref{b}).
\end{enumerate}
\end{definition}
\begin{remark}
\begin{enumerate}[(1)]
\item Notice that if we write
\begin{equation*}
df = \sum_{i=1}^{n}dx^{i}\centerdot\partial_{i}f,
\end{equation*}
then the T-derivative consistency implies 
\begin{align*}
\forall f\in I:\quad \partial_{i}f\in I,\; i=1, \ldots ,n &\Longleftrightarrow \forall f\in I,\;\forall i=1, \ldots ,n:\quad \partial_{i}f = 0\;in\; A = \bigslant{\mathcal{F}}{I},\\
&\Longleftrightarrow \forall f\in I,\; \forall i=1, \ldots ,n:\quad df = 0\;in\; A= \bigslant{\mathcal{F}}{I},\\
&\Longleftrightarrow d(I) = 0\; in\; A.
\end{align*}
\item Because $M_{n}(\mathcal{F})\cong \mathcal{F}\otimes_{\mathbb{F}} M_{n}(\mathbb{F})$ and $M_{n}(A)\cong A\otimes_{\mathbb{F}}M_{n}(\mathbb{F})$, the diagram (\ref{qe}) can be given as
\begin{equation*}
\begin{tikzpicture}
  \node (A) {$\mathcal{F}$};
  \node (B) [below=of A] {$A$};
  \node (AS) [right=of A] {$\mathcal{F}\otimes_{\mathbb{F}}M_{n}(\mathbb{F}) = M_{n}(\mathcal(F))$};
  \node (D) [right=of B] {$\quad A\otimes_{\mathbb{F}}M_{n}(\mathbb{F}) = M_{n}(A) $};
  \draw[-stealth] (A)-- node[left] {\small $\eta_{_{I}}$} (B);
  \draw[-stealth] (B)-- node [below] {\small $\tilde{T}$} (D);
  \draw[-stealth] (A)-- node [above] {\small $T$} (AS);
  \draw[-stealth] (AS)-- node [right] {\small $\eta_{_{I}}\otimes id = M_{n}(\eta_{_{I}})$} (D);
\end{tikzpicture}
\end{equation*}
where $Ker(\eta_{_{I}}\otimes id) = Ker\;\eta_{_{I}}\otimes id =M_{n}(I)$, and where $\eta_{_{I}}\otimes id = M_{n}(\eta_{_{I}})$.
\end{enumerate}
\end{remark}
We can now state the following result.
\begin{result}
An ideal $I\vartriangleleft \mathcal{F}$ is supported by $T:\mathcal{F}\longrightarrow M_{n}(\mathcal{F})$ iff it is both T-Consistent and T-derivative Consistent.
\end{result}
\begin{proof-non}
Suppose $I\vartriangleleft\mathcal{F}$ is supported by $T$; so $A= \bigslant{\mathcal{F}}{I}$ has a FOCDC given by the relation (\ref{b}). Then as we have seen $T$ determines a homomorphism $A\longrightarrow M_{n}(A)$, and this implies $T(I)\subseteq M_{n}(I)$ or equivalently $T^{i}_{j}(I)\subseteq I$. So $I$ is T-Consistent. Moreover, because $d: A\longrightarrow M$ is linear, $d(0) = 0$ holds. Using this fact, if $f\in I$, then $f=0$ in $A= \bigslant{\mathcal{F}}{I}$ and we can write
\begin{equation*}
\forall f\in I\vartriangleleft\mathcal{F}:\quad df = dx^{i}\centerdot\partial_{i}f = 0\Longleftrightarrow\partial_{i}f = 0\; in\; A,\; i= 1, \ldots ,n\Longleftrightarrow \partial_{i}f\in I\;in\;\mathcal{F}\Longleftrightarrow \partial_{i}f\subseteq I.
\end{equation*} 
where the freeness on the right in $\Omega_{d}^{1}(A)$ has been used. This shows that $I$ is T-derivative Consistent.\vspace{.25cm}\\
Conversely, suppose $I\vartriangleleft\mathcal{F}$ is T-Consistent and T-derivative Consistent. By T-Consistency we obtain an induced algebra homomorphism $T:A\longrightarrow M_{n}(A)$ in passing to the quotients. This homomorphism defines a bimodule structure by the relation (\ref{b}). Because $T$ is an algebra homomorphism, the mapping $d: A\longrightarrow A\centerdot d(A)\centerdot A$ satisfies the Leibniz rule. However, this mapping, to be a derivation, must be linear. (this is possible since $d(0) = 0$.) Next, the T-derivative Consistency of $I$ implies $\partial (I)\subseteq I$, or equivalently, $\partial(\bar{0}) = \bar{0}$ in $A$. \\
However, this condition, on uzing $df = \sum_{i=1}^{n}dx^{i}\centerdot\partial_{i}f$, implies that $d(I) = (0)$ (or $d(\bar{0}) = 0\in M_{n}(A)$), and this allows us to assume that $d$ is linear. \\
Therefore, the bimodule structure of the relation (\ref{b}), with this differential mapping specifies a FOCDC over $A$ and $I\vartriangleleft \mathcal{F}$ is supported by $T$.
\end{proof-non}
\begin{definition}
A homomorphism $T: \mathcal{F}\longrightarrow M_{n}(\mathcal{F})$ is called a \texttt{homogeneous homomorphism}, if it acts linearly on the generators of $\mathcal{F}$. Given a homogeneous homomorphism $T$, the corresponding FOCDC over $A= \bigslant{\mathcal{F}}{I}$, is called a \texttt{homogeneous FOCDC over $A$}. 
\end{definition}
Let $T: \mathcal{F}\longrightarrow M_{n}(\mathcal{F})$ be a homogeneous homomorphism and suppose $I(T)\vartriangleleft \mathcal{F}$ is the largest ideal in $\mathcal{F}$ that is supported by $T$. (such an ideal is simply the sum of all ideals in $\mathcal{F}$ that are supported by $T$.) Then the quotient algebra 
\begin{equation}
A(T) := \bigslant{\mathcal{F}}{I(T)}
\end{equation}
is the smallest algebra which has a FOCDC determined by $T$. $A(T)$ is called \texttt{optimal algebra} that has a FOCDC determined by $T$ (as given by the relation (\ref{b})). \\
It is easily verified that a sufficient condition for optimacy of $A$ is given by
\begin{equation}
\forall f\in A:\quad df = 0\Longrightarrow f= constant\;(i.e.\;f \in \mathbb{F}).
\end{equation}
We shall comeback to this concept in the next section. \\
In this final part of this section we shall determine some FOCDC over algebras of interest.
\begin{enumerate}[\underline{\textbf{Case}}~1.]
\item \subsubsection*{Free algebra on 2-generators} \label{wf}
Let $\mathcal{F} = \mathbb{F}\langle x^{1}, x^{2}\rangle$. We want to determine the homogeneous FOCD Calculi on $\mathcal{F}$. Because we are interested in the homogeneous case, the homomorphism $T: \mathcal{F}\longrightarrow M_{2}(\mathcal{F})$ must act linearly on the generators $x^{1}, x^{2}$. Therefore, we may assume that
\begin{equation}
\left\lbrace\begin{array}{rl}T(x^{1})&= Ax^{1} + Bx^{2}\\
T(x^{2})&= Cx^{1} + Dx^{2}
\end{array}\right. \label{qr}
\end{equation}
where $A, B, C, D\in M_{2}(\mathbb{F})$. Note that the relation (\ref{qr}) can be collectively be written as
\begin{equation}
T(x^{1}, x^{2}) = (x^{1}, x^{2})\left( \begin{array}{cc}A& C\\
B & D
\end{array}\right) .
\end{equation}
Let us assume that
\begin{equation*}
A= \left( \begin{array}{cc}a_{11}& a_{12}\\
a_{21} & a_{22}
\end{array}\right) ,\; B=\left( \begin{array}{cc}b_{11}& b_{12}\\
b_{21} & b_{22}
\end{array}\right) ,\; C=\left( \begin{array}{cc}c_{11}& c_{12}\\
c_{21} & c_{22}
\end{array}\right) ,\; D= \left( \begin{array}{cc}d_{11}& d_{12}\\
d_{21} & d_{22}
\end{array}\right) 
\end{equation*}
and also write the relation (\ref{qr}) in the form
\begin{equation}
\left\lbrace\begin{array}{rl}T(x^{1})&= T_{1}^{1}x^{1} + T^{1}_{2}x^{2} =: T^{1}\\
T(x^{2}) &= T^{2}_{1}(x^{1}) + T^{2}_{2}x^{2}=: T^{2}
\end{array}\right. \label{qu}
\end{equation}
$T^{1}$ and $T^{2}$ are $2\times 2$ matrices with entries in $lin_{\mathbb{F}}\lbrace x^{1}, x^{2}\rbrace$. Where $T_{1}^{1} = A,\; T^{1}_{2} = B, \ldots$. Using these notations, the commutation relations
\begin{equation}
x^{j}\centerdot dx^{i} = dx^{k}\centerdot T^{i}_{k}(x^{j}). \label{qi}
\end{equation}
can be written more explicity as
\begin{equation}
x^{j}\centerdot dx^{i} = dx^{k}\centerdot t^{j\;i}_{l\,k}x^{l}. \label{qo}
\end{equation}
where the components $t^{j\;i}_{l\;k}$ are
\begin{equation}
t^{j\;i}_{l\;k} = (T^{i}_{k})^{j}_{l}. \label{qp}
\end{equation}
where the $(j, l)$ indices are those appearing in the relation (\ref{qu}) and $(i, k)$ indices are the column and row indices of the $2\times 2$ matrix $T^{j}_{l}$ appearing in this relation. 
\begin{remark}
In the formula (\ref{qi}), $T^{j\;i}_{\;\;\;k} := T^{i}_{k}(x^{j})$ is a linear form in the coordinates $x^{1}, x^{2}$, i.e. 
\begin{equation*}
T^{j\;i}_{\;\;\;k} := T^{i}_{k}(x^{j}) = t^{j\;i}_{1\;k}x^{1} + t^{j\;i}_{2\;k}x^{2}.
\end{equation*}
\end{remark}
Using the relation (\ref{qo}) and (\ref{qp}), we can now write 
\begin{align}
\nonumber\bullet\qquad\qquad x^{1}\centerdot dx^{1} &= dx^{1}\centerdot (t^{1\;1}_{1\;1}x^{1} + t^{1\;1}_{2\;1}x^{2}) + dx^{2}\centerdot (t^{1\;1}_{1\;2}x^{1} + t^{1\;1}_{2\;2}x^{2}) \\
&= dx^{1}\centerdot (a_{11}x^{1} + b_{11}x^{2}) + dx^{2}\centerdot (a_{21}x^{1} +b_{21}x^{2}).\\
\nonumber\bullet\qquad\qquad x^{1}\centerdot dx^{2} &= dx^{1}\centerdot (t^{1\;2}_{1\;1}x^{1} + t^{1\;2}_{2\;1}x^{2}) + dx^{2}\centerdot (t^{1\;2}_{1\;2}x^{1} + t^{1\;2}_{2\;2}x^{2}),\\
&= dx^{1}\centerdot (a_{12}x^{1} + b_{12}x^{2}) + dx^{2}\centerdot (a_{22}x^{1} + b_{22}x^{2}).\\
\nonumber \bullet\qquad\qquad x^{2}\centerdot dx^{1} &= dx^{1}\centerdot (t^{2\;1}_{1\;1}x^{1} + t^{2\;1}_{2\;1}x^{2}) + dx^{2}\centerdot (t^{2\;1}_{1\;2}x^{1} + t^{2\;1}_{2\;2}x^{2}),\\
&= dx^{1}\centerdot (c_{11}x^{1} + d_{11}x^{2}) + dx^{2}\centerdot (c_{21}x^{1} + d_{21}x^{2}).\\
\nonumber \bullet\qquad\qquad x^{2}\centerdot dx^{2} &= dx^{1}\centerdot (t^{2\;2}_{1\;1}x^{1} + t^{2\;2}_{2\;1}x^{2}) + dx^{2}\centerdot (t^{2\;2}_{1\;2}x^{1} + t^{2\;2}_{2\;2}x^{2}),\\
&= dx^{1}\centerdot (c_{12}x^{1} + d_{12}x^{2}) + dx^{2}\centerdot (c_{22}x^{1} + d_{22}x^{2}).
\end{align}
Summing up, we have the following commutation rules among the coordinates and their differentials
\begin{equation}\label{ry} 
\boxed{
  \!\begin{aligned}
 x^{1}\centerdot dx^{1} &= dx^{1}\centerdot (a_{11}x^{1} + b_{11}x^{2}) + dx^{2}\centerdot (a_{21}x^{1} + b_{21}x^{2}) \\
 x^{1}\centerdot dx^{2} &= dx^{1}\centerdot (a_{12}x^{1} + b_{12}x^{2}) + dx^{2}\centerdot (a_{22} x^{1} + b_{22}x^{2})  \\
 x^{2}\centerdot dx^{1} &= dx^{1}\centerdot (c_{11}x^{1} + d_{11}x^{2}) + dx^{2}\centerdot (c_{21}x^{1} + d_{21}x^{2})  \\
 x^{2}\centerdot dx^{2} &= dx^{1}\centerdot (c_{12}x^{1} + d_{12}x^{2}) + dx^{2}\centerdot (c_{22}x^{1} + d_{22}x^{2})\\
  \end{aligned}
}
\end{equation}
Notice that the communication relations (\ref{ry}) can be compactly written in the form
\begin{equation}
(x^{1}\centerdot dx^{1}, x^{1}\centerdot dx^{2}, x^{2}\centerdot dx^{1}, x^{2}\centerdot dx^{2}) = (dx^{1}\centerdot x^{1}, dx^{2}\centerdot x^{1}, dx^{1}\centerdot x^{2}, dx^{2}\centerdot x^{2})\left( \begin{array}{cc}A& C\\
B & D
\end{array}\right) \label{wi}
\end{equation}
For any arbitrary choice of $A, B, C, D\in M_{2}(\mathbb{F})$, there exists a homogeneous FOCDC on $\mathcal{F} = \mathbb{F}\langle x^{1}, x^{2}\rangle$, with the communication rules  among coordinates and their differentials given by the relations (\ref{ry}). These relations and those coming from the Leibniz rules define the bimodule $\Omega_{d}^{1}(\mathcal{F})$, a FOCDC which is also homogeneous.\\
When we go over to $A= \bigslant{\mathcal{F}}{I}$ the restrictions on $I$:
\begin{enumerate}[(i)]
\item T-Consistency: $T^{i}_{k}(I)\subseteq I$ or $T(I)\subseteq M_{n}(I)$;
\item T-derivative Consistency: $d(I) = 0$ (or $\partial (I) = 0$)
\end{enumerate}
will impose restrictions on the matrices $A, B, C, D$ and consequently the commutation rules (\ref{ry}) will be affected. We shall now consider two important cases to demonstrate how things work.
\item \subsubsection*{2-generated Grassman algebra}
\begin{equation}
\mathcal{A} =\bigslant{ \mathbb{F}\langle x^{1}, x^{2}\rangle}{I},\quad I= ((x^{1})^{2}, (x^{2})^{2}, x^{1}x^{2} + x^{2}x^{1}).
\end{equation}
We shall prove that each homogeneous FOCDC on the Grassmann algebra $\mathcal{A}$ is defined by commutation rules whose related homomorphism $T: \mathcal{A}\longrightarrow M_{2}(\mathcal{A})$ has the form
\begin{equation}
\left\lbrace\begin{array}{rl}T(x^{1}) &= \left( \begin{array}{cc}-1& -c_{11}\\
0 & -1-c_{21}
\end{array}\right)x^{1} + \left( \begin{array}{cc}0& b_{12}\\
0 & b_{22}
\end{array}\right)x^{2}\\
T(x^{2})&= \left( \begin{array}{cc}c_{11}& 0\\
c_{21}& 0
\end{array}\right)x^{1} + \left( \begin{array}{cc}-1-b_{12}& 0\\
-b_{22} & -1
\end{array}\right)x^{2} 
\end{array}\right.
\end{equation}
where $c_{11}, c_{21}, b_{12}, b_{22}\in\mathbb{F}$ are such that
\begin{equation}
det\left( \begin{array}{cc}c_{11}& b_{12}\\
c_{21} & b_{22}
\end{array}\right) = 0.
\end{equation}
\begin{proof-non}
\begin{enumerate}[(i)]
\item Let us first write out the condition of T-derivative Consistency for the relation ideal $I$:
\begin{align*}
d(x^{1})^{2} &= dx^{1}\centerdot x^{1} + x^{1}\centerdot dx^{1} = dx^{1}\centerdot\underbrace{\left\lbrace (a_{11} +1)x^{1} +b_{11}x^{2}\right\rbrace }_{\partial_{1}(x^{1})^{2}} + dx^{2}\centerdot\underbrace{\left\lbrace a_{21} x^{1} +b_{21}x^{2}\right\rbrace}_{\partial_{2}(x^{1})^{2}}  \\
 &= 0
\end{align*}
where the relation (\ref{ry}) has been used in the first line, since we require $\partial_{1}(x^{2}) = 0 = \partial_{2}(x^{1})^{2}$. So we have
\begin{equation*}
a_{11} = -1,\; b_{11} = 0,\; a_{21} = 0,\; b_{21} = 0.
\end{equation*} 
Similarly, we can write using the relatin (\ref{ry}),
\begin{align*}
d(x^{2})^{2} &= dx^{2}\centerdot x^{2} + x^{2}\centerdot dx^{2} = dx^{1}\centerdot\underbrace{\lbrace c_{12}x^{1} + d_{12}x^{2}\rbrace}_{=\partial_{1}(x^{2})^{2}} + dx^{2}\centerdot\underbrace{\lbrace c_{22}x^{1} + (d_{22} +1)c_{22}x^{2}\rbrace}_{=\partial_{2}(x^{2})^{2}},\\
&= 0 \Longrightarrow d_{22} = -1,\; c_{12} = d_{12} = c_{22} = 0
\end{align*}
Finally,
\begin{align*}
d(x^{1}x^{2} + x^{2}x^{1}) &= dx^{1}\centerdot x^{2} + x^{1}\centerdot dx^{2} + dx^{2}\centerdot x^{1} + x^{2}\centerdot dx^{1},\\
&= dx^{1}\centerdot\lbrace(a_{12} + c_{11})x^{1} + (d_{11} + b_{12} + 1)x^{2}\rbrace +\\
&+ dx^{2}\centerdot\lbrace (a_{22} + c_{21} +1)x^{1}+ (d_{21} + b_{22})x^{2}\rbrace =0\\
&\Longrightarrow \left\lbrace\begin{array}{rl}a_{12} + c_{11} = 0&,\;\; d_{11} + b_{12} + 1 =0\\
a_{22} + c_{21} + 1 =0&,\;\; d_{21} + b_{22} = 0
\end{array}\right.
\end{align*}
Thus all together we have obtained the following constraints among the parameters
\begin{equation}
\left\lbrace\begin{array}{rl}b_{11}= a_{21} &= b_{21} = c_{12} = d_{12} = c_{22} = 0\\
a_{11} = d_{22} &= -1\\
a_{12} + c_{11} &= d_{21} + b_{22} = 0\\
a_{22} + c_{21} &= d_{11} + b_{12} = -1
\end{array}\right.
\end{equation}
We can, therefore, choose $c_{11}, b_{22}, b_{12}, c_{21}$ as the independent parameters and obtain
\begin{equation}
A= \left( \begin{array}{cc}-1& -c_{11}\\
0 & -1-c_{21}
\end{array}\right),\; B= \left( \begin{array}{cc}0& b_{12}\\
0 & b_{22}
\end{array}\right),\; C=\left( \begin{array}{cc}c_{11}& 0\\
c_{21} & 0
\end{array}\right),\; D = \left( \begin{array}{cc}-1- b_{12} & 0\\
-b_{22}& -1
\end{array}\right)\label{qa}
\end{equation}
\item We shall next write out the condition of T-Consistency for $I$. In what follows $"\equiv"$ means equality modulo $M_{2}(I)$. 
\begin{align}
\nonumber 0 \equiv T(x^{1})^{2} &= (T(x^{1}))^{2},\quad since\; T\; is\;an\; algebra\; homomorphism.\\
\nonumber &= (Ax^{1} + Bx^{2})^{2} = A^{2}(x^{1})^{2} + B^{2}(x^{2})^{2} + ABx^{1}x^{2} + BAx^{2}x^{1},\\
&= (AB - BA)x^{1}x^{2}\Longrightarrow AB = BA. \label{we}
\end{align}
Similarly, from $T(x^{2})^{2}\equiv 0$ we obtain
\begin{equation}
CD = DC. \label{wq}
\end{equation}
and finally from above relations we obtain
\begin{align}
\nonumber 0\equiv T(x^{1}x^{2} + x^{2}x^{1}) &= T(x^{1})T(x^{2}) + T(x^{2})T(x^{1}),\\
\nonumber &= (Ax^{1} + Bx^{2})(Cx^{1} + Dx^{2}) + (Cx^{1} + Dx^{2})(Ax^{1} + Bx^{2}),\\
\nonumber &= (AC + CA)(x^{1})^{2} + (BD + DB)(x^{2})^{2} + (AD - DA - BC + CB)x^{1}x^{2}\\
\Longrightarrow AD - DA &= BC - CB. \label{wr}
\end{align}
Using the matrices $A, B, C$ and $D$ as given by the relation (\ref{qa}) in the \crefrange{we}{wr}, We compute 
\begin{align*}
AB &= \left( \begin{array}{cc}0& -b_{12}-c_{11}b_{22}\\
0 & -(1+ c_{21})b_{22}
\end{array}\right)  = \left(\begin{array}{cc}0& -(1+ c_{21})b_{12}\\
0 & -(1+ c_{21}b_{22}) 
\end{array}\right) = BA\Longrightarrow c_{11}b_{22} = c_{21}b_{12}.\\
CD &= \left(\begin{array}{cc}-c_{11}(1+ b_{12})& 0\\
-c_{21}(1+ b_{12}) & 0
\end{array}\right) = \left(\begin{array}{cc}-c_{11}(1+ b_{12})& 0\\
-c_{21}-c_{11}b_{22} & 0
\end{array}\right) = DC \Longrightarrow c_{11}b_{22} = b_{12}c_{21}.\\
AD - DA &= \left(\begin{array}{cc}c_{11}b_{22}& -c_{11}b_{12}\\
c_{21}b_{22} & -c_{11}b_{22}
\end{array}\right) = \left(\begin{array}{cc}c_{21}b_{12}& -c_{11}b_{12}\\
c_{21}b_{22}& -c_{21}b_{12}
\end{array}\right) = BC - CB\Longrightarrow c_{11}b_{22} = c_{21}b_{12}.
\end{align*}
We notice that all these three restrictions on the four remaining parameters $c_{11}, b_{22}, c_{21}, b_{12}$ are equal and is in fact the condition
\begin{equation*}
det \left(\begin{array}{cc}c_{11}& b_{12}\\
c_{21} & b_{22}
\end{array}\right) = 0.
\end{equation*}
\end{enumerate}
\end{proof-non} 
Finally we demonstrate that $\mathcal{A}$ is the optimal algebra for every homogenous FOCDC defined on it. To this end, notice that as a linear space over $\mathbb{F}$, $\mathcal{A}$ is spanned by the elements $\lbrace 1, x^{1}, x^{2}, x^{1}x^{2}\rbrace$. If there is an ideal $J\vartriangleleft \mathcal{A}$ such that $I\subset J$, then $J$ must contain the element $x^{1}x^{2}\in\mathcal{A}$ and consequently the condition $d(x^{1}x^{2}) = 0$ must hold. We show that this condition can not be satisfied!
\begin{align*}
0 &= d(x^{1}x^{2}) = dx^{1}\centerdot x^{2} + x^{1}\centerdot dx^{2},\\
&= dx^{1}\centerdot\lbrace a_{12}x^{1} + (1+ b_{12})x^{2}\rbrace + dx^{2}\centerdot\lbrace a_{22}x^{1} + b_{22}x^{2}\rbrace ,\\
&= dx^{1}\centerdot\lbrace -c_{11}x^{1} + (1+ b_{12})x^{2}\rbrace + dx^{2}\centerdot\lbrace -(1+ c_{21})x^{1} + b_{22}x^{2}\rbrace .\\
&\Longrightarrow c_{11} = 0 = b_{22},\qquad b_{12} = -1 = c_{21}.
\end{align*}
However, this implies $det\left(\begin{array}{cc}c_{11}& b_{12}\\
c_{21} & b_{22}
\end{array}\right) = -1 \neq 0
$, which contradicts the result obtained earlier, namely this determinant must be zero.\\
This proves the required optimacy for $\mathcal{A}$.
\item \subsubsection*{Quantum Plane $H_{2}$}
\begin{equation*}
H_{2} = \bigslant{\mathbb{C}\langle x^{1}, x^{2}\rangle}{(x^{1}x^{2} - q^{-1}x^{2}x^{1})},\quad q\neq 1.
\end{equation*}
It turns out that a variety of coordinate diff. calculi on $H_{2}$ crucially depends on $q$ being or not equal to $-1$.\\
Just as in the Grassmann case, we start off with applying T- derivative invariance for the relation ideal $I = (x^{1}x^{2} - q^{-1}x^{2}x^{1})$. Using the relation (\ref{ry}) we compute:
\begin{align*}
0 &= d(x^{1}x^{2} - q^{-1}x^{2}x^{1}) = dx^{1}\centerdot x^{2} + x^{1}\centerdot dx^{2} - q^{-1}dx^{2}\centerdot x^{1} - q^{-1}x^{2}\centerdot dx^{1} \\
&= dx^{1}\centerdot\lbrace x^{1}(a_{12} - q^{-1}c_{11}) + x^{2}(1+ b_{12} - q^{-1}d_{11})\rbrace +\\
&+ dx^{2}\centerdot\lbrace x^{1}(a_{22} - q^{-1} - q^{-1}c_{21}) + x^{2}(b_{22} - q^{-1}d_{21})\rbrace .
\end{align*}
This shows that $I$ is T-derivative invariant iff
\begin{equation}
\left\lbrace\begin{array}{rl}qa_{12} - c_{11} = 0,&\quad q(1+b_{12}) = d_{11}\\
qa_{22} = 1+ c_{21},&\quad d_{21} = qb_{22}
\end{array}\right. \label{wy}
\end{equation}
Next, we apply the condition of T-invariance,
\begin{align}
\nonumber 0 &\equiv T(x^{1}x^{2} -q^{-1}x^{2}x^{1}) = T(x^{1})T(x^{2}) - q^{-1}T(x^{2})T(x^{1}),\\
\nonumber &= (Ax^{1} + Bx^{2})(Cx^{1} + Dx^{2}) - q^{-1}(Cx^{1} + Dx^{2})(Ax^{1} + Bx^{2}),\\
\nonumber &= (AC - q^{-1}CA)(x^{1})^{2} + (BD - q^{-1}DB)(x^{2})^{2} + (AD - DA + qBC - q^{-1}CB)x^{1}x^{2}.\\
&\Longrightarrow\left\lbrace\begin{array}{rl}AC - q^{-1}CA &= 0\\
BD - q^{-1}BD &= 0
\end{array}\right. \label{wu}
\end{align}
The conditions (\ref{wy}) and (\ref{wu}) can be solved to determine the matrices $A, B, C$ and $D$. One obtains several classes of solutions as given by \cite{ulya}
\begin{remark}
\begin{enumerate}[(1)]
\item The algebra $H_{2} = \bigslant{\mathbb{C}\langle x^{1}, x^{2}\rangle}{(x^{1}x^{2} - q^{-1}x^{2}x^{1})}$ has an automorphism group given by\\
(a) If $q^{2}\neq 1$, $ Aut(H_{2})\cong (\mathbb{C}^{\ast})^{2}$  for Torus $\mathbb{C}^{\ast} = \mathbb{C} - \lbrace 0\rbrace $. which acts naturally on $\mathbb{C}x^{1}\oplus\mathbb{C}x^{2}$.\\
(b) If $q= -1$, $Aut(H_{2})$ is isomorphic to the Torus $(\mathbb{C}^{\ast})^{2}$ and the symmetry exchanging $x^{1}$ and $x^{2}$.\vspace*{.5cm}\\
In what we have said above, we are interested in the case (a). It can be shown that in this case, the automorphism of $H_{2}$ corresponding to 
\begin{equation*}
x^{1}\longrightarrow x^{1},\quad x^{2}\longrightarrow \alpha x^{2},\; \alpha\in\mathbb{C}^{\ast}.
\end{equation*}
transforms the commutation rules given by
\begin{equation*}
\left(\begin{array}{cc}A&C\\
B & D
\end{array}\right) = \left(\begin{array}{ccccc}a_{11}& a_{12} & \vdots & c_{11} & c_{12}\\
a_{21} & a_{22} &\vdots & c_{21} & c_{22}\\
\cdots &\cdots &\cdots &\cdots & \cdots \\
b_{11} & b_{12} &\vdots & d_{11} & d_{12}\\
b_{21} & b_{22} & \vdots & d_{21} & d_{22}
\end{array}\right)
\end{equation*}
as specified by the relation (\ref{wi}) into
\begin{equation*}
\left(\begin{array}{ccccc}a_{11}& \alpha a_{112} &\vdots & \alpha c_{11} &\alpha^{2}c_{12}\\
\alpha a_{21}& a_{22} & \vdots & c_{21} & \alpha c_{22}\\
\cdots &\cdots &\cdots &\cdots &\cdots \\
\alpha^{-1}b_{11} & b_{12} &\vdots & d_{11} &\alpha d_{12}\\
\alpha^{-2}b_{21} & \alpha^{-1}b_{22} &\vdots & \alpha^{-1}d_{21} & d_{22}
\end{array}\right)
\end{equation*}
such transformations do not mix up distinct set of solutions $\lbrace A, B, C, D\rbrace$.
\item It can be shown that on the quantum hyperplane (for $n\geq 3$)
\begin{equation}
H_{n} = \bigslant{\mathbb{C}\langle x^{1}, \ldots , x^{n}\rangle}{(x^{i}x^{j} - q^{-1}x^{j}x^{i}\vert i< j)} \label{wa}
\end{equation}
there exists only one FOCDC (up to the exchange of $q$ with $q^{-1}$) which is given by
\begin{equation}\label{wp} 
\boxed{
  \!\begin{aligned}
x^{i}\centerdot dx^{j} = \left\lbrace\begin{array}{rl}qdx^{j}\centerdot x^{i}\qquad &for\; i<j\\
q^{2}dx^{j}\centerdot x^{i}\qquad &for \;i=j\\
qdx^{j}\centerdot x^{i} + (q^{2} - 1)dx^{i}\centerdot dx^{j} \quad &for\; i>j
\end{array}\right.
  \end{aligned}
}
\end{equation}
This is called the \texttt{Pusz- Woronowicz calculus}
\end{enumerate}
\end{remark}
About the optimacy of the algebra (\ref{wa}) the following result has been obtained:\vspace*{.5cm}\\
Let $\lbrace q\rbrace$ denote the minimal positive natural numbers $m$ such that $q^{2m} = 1$. If no such number exists, we write $\lbrace q\rbrace = 0$.
\begin{theorem}
For P-W calculus (\ref{wp}), the optimal algebra is
\begin{enumerate}[(i)]
\item $H_{n}$, as given by the relation (\ref{wa}), if $\lbrace q\rbrace \leqslant 1$.
\item If $\lbrace q\rbrace = m>1$, the optimal algebra is 
\begin{equation*}
\bigslant{H_{n}}{\left( (x^{1})^{m},\ldots , (x^{n})^{m}\right) }.
\end{equation*}
\end{enumerate}
\end{theorem}
\end{enumerate}
\section{Optimal algebras for FOCD Calculi}
As before let $\mathcal{F} = \mathbb{F}\langle x^{1}, \ldots , x^{n}\rangle$ and $T: \mathcal{F}\longrightarrow M_{n}(\mathcal{F})$ be an algebra homomorphism. Recall that a 2-sided ideal $I\vartriangleleft\mathcal{F}$ is said to be T-Consistent (or T-invariant) if $T^{i}_{k}(I)\subseteq I$. An ideal $I\vartriangleleft\mathcal{F} $ is said to be T-derivative Consistent (or T-derivative invariant ) if $\partial_{k}(I)\subseteq I$ for every partial derivative $\partial_{k}, k=1, \ldots ,n$, defined by the differential $d$ corresponding to $T$. (This is the differential mapping of the algebra $\mathcal{F},\; d: \mathcal{F}\longrightarrow \mathcal{F}\centerdot d(\mathcal{F})\centerdot \mathcal{F}$. It is called a \texttt{Cover differential} for any quotient algebra $A= \bigslant{\mathcal{F}}{I}$). \\
Given the homomorphism $T$, there exists the largest T-Consistent and T-derivative Consistent ideal $I(T)\vartriangleleft \mathcal{F}$, (which is the sum of all ideals which are T-Consistent and T-derivative Consistent). 
\begin{definition}
Let $I(T)$ be the sum of all T-Consistent and T-derivative Consistent ideals in $\mathcal{F}$. The factor algebra
\begin{equation}
A(T) := \bigslant{\mathcal{F}}{I(T)}.
\end{equation}
is said to be the optimal algebra for the FOCDC given by the commutation rule
\begin{equation}
x^{i}\centerdot dx^{j} = dx^{k}\centerdot T^{i}_{k}(x^{j}),\quad i,j = 1, \ldots ,n.
\end{equation}
A triple $\left( A(T), d, \Omega^{1}_{d}(A(T)))\right) $, where $d$ is the cover differential, is called an optimal calculus.
\end{definition}
We have seen that the free algebra $\mathcal{F}$ admits a FOCDC for arbitrary commutation rules (i.e for arbitrary homomorphism $T: \mathcal{F}\longrightarrow M_{n}(\mathcal{F}) $). In order to define the homomorphism $T$, it is enough to set its values on generators via 
\begin{equation}
T^{j}_{k}(x^{i}) = t^{i\;j}_{k} + t^{i\;j}_{k\;l}x^{l} + t^{i\;j}_{k\;l_{1}\;l_{2}}x^{l_{1}}x^{l_{2}} + \cdots . \label{ws}
\end{equation}
where $\lbrace t^{i\;j}_{k\;l_{1}\;l_{2}\cdots}\rbrace$ are arbitrary tensor coefficients. If we require the homomorphism $T$ to preserve degree (in which case it is called a homogeneous homomorphism) it must acts linearly on the generators $x^{i}, i=1, \ldots ,n$ of $\mathcal{F}$, that is
\begin{equation}
T^{j}_{k}(x^{i}) = t^{i\;j}_{k\;l}x^{l}.\quad summation\;over\;l. \label{wj}
\end{equation} 
The general case has been considered in \cite{khar}\\
We shall here consider the homogeneous case (\ref{wj}) and determine the optimal algebras for such homogeneous FOCDCalculi. It is seen from the relation (\ref{wa}) that a homogeneous homomorphism $T$ is determined by a 2-Covariant 2-Contravariant tensor $T = (t^{i\;j}_{k\;l})$. Using the relation (\ref{ws}) in (\ref{wj}) we can write the commutation rule of such calculi as
\begin{equation}
x^{i}\centerdot dx^{j} = dx^{k}\centerdot T^{j}_{k}(x^{i}) = dx^{k}\centerdot t^{j\;i}_{l\;k}x^{l}. \label{wv}
\end{equation} 
In this notation
\begin{equation}
T^{j}_{k}(x^{i}) =: T^{j\;i}_{\;\;\;k}. \label{wd}
\end{equation}
where $T^{j}$ is an $n\times n$ matrix in $M_{n}(\mathcal{F})$, whose (i-k) entries is given by the relation (\ref{wd}). The entries of $T^{j}$ are linear in $x^{1}, \ldots , x^{n}$, i.e. $T^{j\;i}_{\;\;\;k}\in lin_{\mathbb{F}}\lbrace x^{1}, \ldots , x^{n}\rbrace$. To make contact with our previous notations notice that
\begin{equation*}
T(x^{j}) := T^{j} := T^{j}_{l}x^{l} = T^{j}_{1}x^{1} + T^{j}_{2}x^{2}+ \cdots .
\end{equation*}  
where in the case of two generators $x^{1}, x^{2}$, this is just the notation of free algebra on 2-generators ( Case \ref{wf}):
\begin{equation*}
T(x^{1}) = T^{1}_{1}x^{1} + T^{1}_{2}x^{2},\qquad T(x^{2}) = T^{2}_{1}x^{1} + T^{2}_{2}x^{2}.
\end{equation*} 
where $T^{1}_{1}, T^{1}_{2}, T^{2}_{1}, T^{2}_{2}\in M_{2}(\mathbb{F})$. Therefore, in the notation (\ref{wd}), $(i, k)$ are the column, row indices of the matrix $T^{j}$. When the tensor $t^{j\;i}_{l\;k}$ is used, one must notice that the indices $(j, l)$ determine different matrices and each $(j, l)$ matrix has $(i, k)$ as its column and row indices, respectively.
\begin{result}\label{re1}
For any 2-Covariant 2-Contravariant tensor $T= \lbrace t^{j\;i}_{l\;k}\rbrace$ the ideal $I(T)$ can be constructed by induction as a homogeneous space
\begin{equation*}
I(T) = I_{1}(T) + I_{2}(T) + I_{3}(T) + \cdots .
\end{equation*}
in the following manner:
\begin{enumerate}[(1)]
\item $I_{1}(T) = 0$,
\item Assume that $I_{s-1}(T)$ has been defined and let $U_{s}$ be the space of all polynomials $f$ of degree $s$ such that
\begin{equation}
\partial_{k}(f)\in I_{s-1}(T),\quad k=1, \ldots ,n.
\end{equation}
Then $I_{s}(T)$ is the largest T-Consistent (i.e. T-invariant) subspace of $U_{s}$. The ideal $I(T)$ is a maximal T-Consistent ideal in $\mathcal{F}$.
\end{enumerate} 
\end{result}
\begin{proof-non}
See Ref (\cite{ulya} and \cite{boro})
\end{proof-non}
This result shows that in particular if a homogeneous element is such that all elements of the invariant subspace generated by it have all partial derivatives equal to zero, that element equals zero in the optimal algebra.\vspace{.5cm}\\
We shall now consider some explicit examples which show how to determine the optimal algebra.
\begin{exmp}\label{ex11}
We show that for the commutation rules given by
\begin{equation}
\left\lbrace\begin{array}{rl}x^{1}\centerdot dx^{1} &= dx^{1}\centerdot\mu x^{2}\\
x^{1}\centerdot dx^{2} &= -dx^{1}\centerdot x^{2}\\
x^{2}\centerdot dx^{1} &= -dx^{2}\centerdot x^{1}\\
x^{2}\centerdot dx^{2} &= dx^{2}\centerdot\lambda x^{1}
\end{array}\right. \label{wk}
\end{equation}
The optimal algebra is $A(T) = \bigslant{\mathbb{F}\langle x^{1}, x^{2}\rangle}{(x^{1}x^{2}, x^{2}x^{1})}$.
\end{exmp}
\begin{proof-non}
Let $I = (x^{1} x^{2}, x^{2}x^{1})\vartriangleleft\mathcal{F} = \mathbb{F}\langle x^{1}, x^{2}\rangle$. Let $d$ be the free differential for $\mathcal{F}$ (which is also called the cover differential). We notice that
\begin{equation*}
d(x^{1}x^{2}) = dx^{1}\centerdot x^{2} + x^{1}\centerdot dx^{2} = dx^{1}\centerdot x^{2}- dx^{1}\centerdot x^{2} = 0.
\end{equation*}
Similarly,
\begin{equation*}
d(x^{2}x^{1}) = dx^{2}\centerdot x^{1} + x^{2}\centerdot dx^{1} = dx^{2}\centerdot x^{1} - dx^{2}\centerdot x^{1} =0.
\end{equation*}
This shows that the ideal $I$ is T-derivative Consistent (where $T$ is hidden in the relations (\ref{wk})).\\
Next, we show that $I$ is T-Consistent. For this purpose we must use the general form of the commutation relations for $T: \mathcal{F}\longrightarrow M_{2}(\mathcal{F})$, 
\begin{equation}
x^{i}\centerdot dx^{j} = dx^{k}\centerdot T^{j}_{i}(x^{i}) = dx^{k}\centerdot t^{i\;j}_{l\;k}x^{l}.\label{wz}
\end{equation}
where $(i, l)$ specify a $2\times 2$ matrix and $(j, k)$ are the column and row indices of such a matrix.\\
We first consider the commutation relations (\ref{wk}) as if they are for $\mathcal{F}$ (we can do this because $\mathcal{F}$ admits any arbitrary calculus) and determine the form of $T$; i.e. we find the matrices $T^{1} := T(x^{1})$ and $T^{2} = T(x^{2})$; $T^{i},\; i= 1, 2$.\\
Using the relations (\ref{wk}), we can write 
\begin{align}
\nonumber x^{1}\centerdot dx^{1} &= dx^{k}\centerdot t^{1\;1}_{l\;k}x^{l} = dx^{1}\centerdot t^{1\;1}_{1\;1}x^{1} + dx^{1}\centerdot t^{1\;1}_{2\;1}x^{2} +dx^{2}\centerdot t^{1\;1}_{1\;2}x^{1} + dx^{2}\centerdot t^{1\;1}_{2\;2}x^{2},\\
\nonumber &= dx^{1}\centerdot\mu x^{2},\\
&\Longrightarrow t^{1\;1}_{1\;1} = t^{1\;1}_{1\;2} = t^{1\;1}_{2\;2} = 0,\quad t^{1\;1}_{2\;1} = \mu \label{wx}
\end{align}
Similarly, using the relation (\ref{wz}) we write:
\begin{align}
\nonumber x^{1}\centerdot dx^{2} &= dx^{k}\centerdot t^{1\;2}_{l\;k}x^{l} = dx^{1}\centerdot t^{1\;2}_{1\;1}x^{1} + dx^{1}\centerdot t^{1\;2}_{2\;1}x^{2} + dx^{2}\centerdot t^{1\;2}_{1\;2}x^{1} + dx^{2}\centerdot t^{1\;2}_{2\;2}x^{2},\\
\nonumber &= -dx^{1}\centerdot x^{2},\\
&\Longrightarrow t^{1\;2}_{1\;1} = t^{1\;2}_{1\;2} = t^{1\;2}_{2\;2} = 0,\quad t^{1\;2}_{2\;1} = -1. \label{wc}
\end{align}
From the relations (\ref{wx}) and (\ref{wc}) we obtain, using the definition 
\begin{equation*}
T^{1} := T(x^{1}) = T^{1}_{1}x^{1} + T^{1}_{2}x^{2}.
\end{equation*}
But
\begin{equation*}
\left( T^{1}_{1}\right) ^{j}_{k} = \left( t^{1\;j}_{1\;k}\right)  = \left(\begin{array}{cc}0& 0\\
0 & 0
\end{array}\right),\quad \left( T^{1}_{2}\right) ^{j}_{k} = \left( t^{1\;j}_{2\;k}\right) = \left(\begin{array}{cc}\mu & -1\\
0 & 0
\end{array}\right)
\end{equation*}
where relations(\ref{wx}) and (\ref{wc}) are used. So
\begin{equation}
T^{1} := T(x^{1}) = 0x^{1} + \left(\begin{array}{cc}\mu & -1\\
0 & 0
\end{array}\right)x^{2} = \left(\begin{array}{cc}\mu x^{2}& -x^{2}\\
0 & 0
\end{array}\right). \label{wn}
\end{equation}
Next we use the relation (\ref{wv}) to write
\begin{align}
\nonumber x^{2}\centerdot dx^{1} &= dx^{k}\centerdot t^{2\;1}_{l\;k}x^{l} = dx^{1}\centerdot t^{2\;1}_{1\;1}x^{1} + dx^{1}\centerdot t^{2\;1}_{2\;1}x^{2} + dx^{2}\centerdot t^{2\;1}_{1\;2}x^{1} + dx^{2}\centerdot t^{2\;1}_{2\;2}x^{2},\\
\nonumber &= -dx^{2}\centerdot x^{1}.\\
&\Longrightarrow t^{2\;1}_{1\;1} = t^{2\;1}_{2\;2} = t^{2\;1}_{2\;1} = 0,\quad t^{2\;1}_{1\;2} = -1.
\end{align}
Similarly, using the relation (\ref{wv}) we write
\begin{align}
\nonumber x^{2}dx^{2} &= dx^{k}\centerdot t^{2\;2}_{l\;k}x^{l} = dx^{1}\centerdot t^{2\;2}_{1\;1}x^{1} + dx^{1}\centerdot t^{2\;2}_{2\;1}x^{2} + dx^{2}\centerdot t^{2\;2}_{1\;2}x^{1} + dx^{2}\centerdot t^{2\;2}_{2\;2}x^{2},\\
\nonumber &= dx^{2}\centerdot\lambda x^{1}.\\
&\Longrightarrow t^{2\;2}_{1\;1} = t^{2\;2}_{2\;1} = t^{2\;2}_{2\;2} = 0,\quad t^{2\;2}_{1\;2} = \lambda .
\end{align}
It follows that 
\begin{align}
\nonumber T^{2} := T(x^{2}) &= T^{2}_{1}x^{1} + T^{2}_{2}x^{2},\qquad\; T^{2}_{2} = \left( t^{2\;j}_{2\;k}\right) =\left(\begin{array}{cc}0& 0\\
0 & 0
\end{array}\right),\\
&= T^{2}_{1}x^{1} +0 =\left(\begin{array}{cc}0& 0\\
-x^{1} & \lambda x^{1}
\end{array}\right). \label{wm}
\end{align}
Using the relations (\ref{wn}) and (\ref{wm}) we can write 
\begin{align*}
T(x^{1}x^{2}) &= T(x^{1})T(x^{2}) = \left(\begin{array}{cc}\mu x^{2}& -x^{2}\\
0 & 0
\end{array}\right)\left(\begin{array}{cc}0& 0\\
-x^{1} &\lambda x^{1}
\end{array}\right)= \left(\begin{array}{cc}x^{2}x^{1}& -\lambda x^{2}x^{1}\\
0 & 0
\end{array}\right)\in M_{2}(I);\\
&\Longrightarrow T(x^{1}x^{2})\equiv 0\;mod(I).
\end{align*}
Similarly,
\begin{align*}
T(x^{2}x^{1}) &= T(x^{2})T(x^{1}) = \left(\begin{array}{cc}0& 0\\
-\mu x^{1}x^{2} & x^{1}x^{2}
\end{array}\right)\in M_{2}(I),\\
&\Longrightarrow T(x^{2}x^{1}) \equiv 0\;mod(I).
\end{align*}
This proves that $I$ is a T-Consistent ideal. We therefore, conclude that $A(T) := \bigslant{\mathcal{F}}{I}$ admits the FOCDC given by the relations (\ref{wk}).\vspace{.5cm}\\
We now consider the question of optimacy of $A(T)$. In the algebra $\bigslant{\mathbb{F}\langle x^{1}, x^{2}\rangle}{(x^{1}x^{2}, x^{2}x^{1})}$ every element has a unique presentation in the form 
\begin{equation}
f = \alpha_{1}x^{1} + \alpha_{2}(x^{1})^{2} + \ldots + \alpha_{n}(x^{1})^{n} + \beta_{1}x^{2} + \beta_{1}(x^{2})^{2} + \ldots + \beta_{m}(x^{2})^{m},\; m,n\in\mathbb{N}. \label{eq10}
\end{equation}
We will show that 
\begin{equation}
\partial_{k}f \equiv 0,\; k= 1, 2\Longrightarrow f=0\;in\; A(T),\;(i.e.\;f\equiv 0\;mod(I)).
\end{equation}
and this, by the result (\ref{re1}), will imply that $A(T)$ is the optimal algebra for the commutation rule (\ref{wk}). We compute:
\begin{align*}
\partial_{2}(x^{1})^{n} &= \partial_{2}[x^{1}\centerdot (x^{1})^{n-1}] = \partial_{2}x^{1}\centerdot (x^{1})^{n-1} + T^{k}_{2}(x^{1})\centerdot \partial_{k}(x^{1})^{n-1},\\
&= T^{k}_{2}(x^{1})\partial_{k}(x^{1})^{n-1},\quad for\;\partial_{2}x^{1} = 0,\\
&= T^{1}_{2}(x^{1})\partial_{1}(x^{1})^{n-1} + T^{2}_{2}(x^{1})\partial_{2}(x^{1})^{n-1}.
\end{align*}
But
\begin{align}
\nonumber T^{1}_{2}(x^{1}) &= t^{1\;1}_{l\;2}x^{l} = t^{1\;1}_{1\;2}x^{1} + t^{1\;1}_{2\;2}x^{2} = 0x^{1} + 0x^{2} = 0,\\
\nonumber T^{2}_{2}(x^{1}) &= t^{1\;2}_{l\;2}x^{l} = t^{1\;2}_{1\;2}x^{1} + t^{1\;2}_{2\;2}x^{2} = 0x^{1} + 0x^{2} = 0,\\
\Longrightarrow \partial_{2}(x^{1})^{n} &= 0.
\end{align}
Similarly we compute,
\begin{align*}
\partial_{1}(x^{2})^{n} &= \partial_{1}[x^{2}\centerdot (x^{2})^{n-1}] = \partial_{1}x^{2}\centerdot (x^{2})^{n-1} + T^{k}_{1}(x^{2})\partial_{k}(x^{2})^{n-1},\\
&= T^{k}_{1}(x^{2})\partial_{k}(x^{2})^{n-1},\qquad for\; \partial_{1}x^{2} = 0,\\
&= T^{1}_{1}(x^{2})\partial_{1}(x^{2})^{n-1} + T^{2}_{1}(x^{2})\partial_{2}(x^{2})^{n-1}.
\end{align*}
But,
\begin{align}
\nonumber T^{1}_{1}(x^{2}) &= t^{2\;1}_{l\;1}x^{l} =t^{2\;1}_{1\;1}x^{1} + t^{2\;1}_{2\;1}x^{2} = 0x^{1} + 0x^{2} = 0,\\
\nonumber T^{2}_{1}(x^{2}) &= t^{2\;2}_{l\;1}x^{l} = t^{2\;2}_{1\;1}x^{1} + t^{2\;2}_{2\;1}x^{2} = 0x^{1} + 0x^{2} = 0.\\
&\Longrightarrow \partial_{1}(x^{2})^{n} = 0.
\end{align}
Next, we compute,
\begin{align*}
\partial_{1}(x^{1})^{n} &= \partial_{1}[x^{1}\centerdot (x^{1})^{n-1}] = \partial_{1}x^{1}\centerdot (x^{1})^{n-1} + T^{k}_{1}(x^{1})\centerdot\partial_{k}(x^{1})^{n-1},\\
&= (x^{1})^{n-1} + T^{1}_{1}(x^{1})\centerdot\partial_{1}(x^{1})^{n-1} + T^{2}_{1}(x^{1})\centerdot\partial_{2}(x^{1})^{n-1},\\
&= (x^{1})^{n-1} + T^{1}_{1}(x^{1})\centerdot\partial_{1}(x^{1})^{n-1},\qquad for\;\partial_{2}(x^{1})^{n-1} = 0.
\end{align*}
But,
\begin{align}
\nonumber T^{1}_{1}(x^{1}) &= t^{1\;1}_{l\;1}x^{l} = t^{1\;1}_{1\;1}x^{1} + t^{1\;1}_{2\;1}x^{2} = 0x^{1} + \mu x^{2} = \mu x^{2},\\
\nonumber T^{2}_{1}(x^{1}) &= t^{1\;2}_{l\;1}x^{l} = t^{1\;2}_{1\;1}x^{1} + t^{1\;2}_{2\;1}x^{2} = 0x^{1} -x^{2} = -x^{2}.\\
\nonumber\therefore\qquad\partial_{1}(x^{1})^{n} &= (x^{1})^{n-1} + \mu x^{2}\centerdot\partial_{1}(x^{1})^{n-1} -x^{2}\centerdot \partial_{2}(x^{1})^{n-1},\\
&= (x^{1})^{n-1} + \mu x^{2}\centerdot\partial_{1}(x^{1})^{n-1}. \label{eq1}
\end{align}
Similarly we can write, using the relation (\ref{eq1}),
\begin{equation*}
\partial_{1}(x^{1})^{n-1} = (x^{1})^{n-2} + \mu x^{2}\centerdot\partial_{1}(x^{1})^{n-2}.
\end{equation*}
which, on substitution in the relation (\ref{eq1}) yields
\begin{align*}
\partial_{1}(x^{1})^{n} &= (x^{1})^{n-1} + \mu x^{2}\centerdot (x^{1})^{n-2} + (\mu x^{2})^{2}\centerdot\partial_{1}(x^{1})^{n-2} = \ldots \\
&= (x^{1})^{n-1} + \mu x^{2}\centerdot (x^{1})^{n-2} + (\mu x^{2})^{2}\centerdot (x^{1})^{n-3} + \cdots + (\mu x^{2})^{n-2}\centerdot x^{1} + (\mu x^{2})^{n-1},\\
&\equiv (x^{1})^{n-1} + (\mu x^{2})^{n-1}\quad mod(I).
\end{align*}
We may now go back to $f\in A(T)$, given by the relation (\ref{eq10}) and compute
\begin{equation*}
\partial_{1}f = \alpha_{1} + \alpha_{2}x^{1} + \ldots + \alpha_{n}(x^{1})^{n-1} + \alpha_{2}\mu x^{2} + \alpha_{3}(\mu x^{2})^{2} + \ldots + \alpha_{n}(\mu x^{2})^{n-1}.
\end{equation*} 
where we used the fact that $\partial_{1}(x^{2})^{n} = 0$.
\begin{equation*}
\Longrightarrow\partial_{1}f \equiv 0\;mod(I)\Longleftrightarrow \alpha_{1} = \alpha_{2}= \ldots = \alpha_{n} = 0.
\end{equation*}
similarly, one may show that
\begin{equation*}
\partial_{2}f\equiv 0\;mod(I)\Longleftrightarrow\beta_{1} = \beta_{2} = \ldots = \beta_{m} = 0.
\end{equation*}
Thus we have shown that
\begin{equation*}
\partial_{k}f \equiv 0\;mod(I),\; k= 1, 2\Longrightarrow f\equiv 0\;mod(I).
\end{equation*}
that is $f\in I$. This proves that $I$ is a maximal ideal in $\mathcal{F}$ that is T-Consistent and T-derivative Consistent. Hence by the result (\ref{re1}), $A(T) := \bigslant{\mathbb{F}\langle x^{1}, x^{2}\rangle}{(x^{1}x^{2}, x^{2}x^{1})}$ is the optimal algebra for the FOCDC on it given explicity by the relation (\ref{wk}).
\end{proof-non}
\begin{exmp}
Consider the diagonal commutation rules 
\begin{equation}
\left\lbrace\begin{array}{rl}x^{j}\centerdot dx^{i} &= dx^{i}\centerdot q^{ij}x^{j}\\
q^{ij}q^{ji} &= 1,\quad i\neq j
\end{array}\right. \label{eq11}
\end{equation}
We will prove that 
\begin{enumerate}[(1)]
\item If non of the coefficients $q^{ii}$ is a root of polynomial $\lambda^{[m]} := \lambda^{m-1} + \lambda^{m-2} + \ldots + 1$, then optimal algebra is 
\begin{equation*}
A(T) = \bigslant{\mathbb{F}\langle x^{1}, \ldots , x^{n}\rangle}{(q^{ij}x^{i}x^{j} - x^{j}x^{i}\vert i< j)}
\end{equation*}
\item If $(q^{ii})^{[m_{i}]} = 0\quad 1\leq i\leq s $, with minimal $m_{i}$, then
\begin{equation*}
A(T) = \bigslant{\mathbb{F}\langle x^{1}, \ldots , x^{n}\rangle}{(q^{ij}x^{i}x^{j} - x^{j}x^{i},\; i<j,\; (x^{i})^{m_{i}},\; 1\leq i\leq s )}.
\end{equation*}
\end{enumerate}
\end{exmp}
\begin{proof-non}
\begin{enumerate}[(1)]
\item Our strategy is the same. We know that the commutation rules (\ref{eq11}) hold for the free algebra $\mathcal{F}$ (since $T: \mathcal{F}\longrightarrow M_{n}(\mathcal{F})$ can be arbitrarily defined by the values of $T$ on the generators). Actually using the relations (\ref{eq11}) we can find $T$ by comparing the relations (\ref{eq11}) with the standard commutation rule
\begin{equation*}
x^{j}\centerdot dx^{i} = dx^{k}T^{i}_{k}(x^{j}).
\end{equation*}
and we immediately get
\begin{equation}
T^{i}_{k}(x^{j}) = \delta^{i}_{k}q^{ij}x^{j}. \label{eq12}
\end{equation}
We must show that the ideal $I= (q^{ij}x^{i}x^{j} - x^{j}x^{i}\vert i<j)$ is T-Consistent (i.e. T-invariant) under the algebra homomorphism. $T: \mathcal{F}\longrightarrow M_{n}(\mathcal{F})$ given by the relation (\ref{eq12}). To this end let $l<j$ so $q^{lj}x^{l}x^{j} - x^{j}x^{l}\in I$, and compute 
\begin{align*}
T^{i}_{k}(q^{lj}x^{l}x^{j} - x^{j}x^{l}) &= q^{lj}T^{i}_{s}(x^{l})T^{s}_{k}(x^{j}) - T^{i}_{s}(x^{j})T^{s}_{k}(x^{l}),\\
&= q^{lj}\delta^{i}_{s}q^{il}x^{l}\centerdot\delta^{s}_{k}q^{sj}x^{j} - \delta^{i}_{s}q^{ij}x^{j}\centerdot\delta^{s}_{k}q^{sl}x^{l},\\
&= \delta^{i}_{k}q^{kl}q^{kj}(q^{lj}x^{l}x^{j} - x^{j}x^{l}) \equiv 0\;mod(I).
\end{align*}
This shows that $I$ is T-Consistent (or T-invariant).\\
Next we use the partial derivatives corresponding to $T$ (i.e. the partial derivatives corresponding to the cover differential of $\mathcal{F}$ defined by $T$) and compute
\begin{align*}
\partial_{k}(q^{lj}x^{l}x^{j} - x^{j}x^{l}) &= q^{lj}\left[ \underbrace{\partial_{k}x^{l}}_{=\delta^{l}_{k}}\centerdot x^{j} +\underbrace{T^{i}_{k}(x^{l})}_{=\delta^{i}_{k}q^{il}x^{l}}\centerdot\underbrace{\partial_{i}x^{j}}_{=\delta^{j}_{i}}\right]  - \left[ \underbrace{\partial_{k}x^{j}}_{=\delta^{j}_{k}}\centerdot x^{l} + \underbrace{T^{i}_{k}(x^{j})}_{=\delta^{i}_{k}q^{ij}x^{j}}\centerdot\underbrace{\partial_{i}x^{l}}_{=\delta^{l}_{i}}
\right] \\
&= q^{lj}\delta^{l}_{k}x^{j} + q^{lj}\delta^{i}_{k}q^{il}x^{l}\delta^{j}_{i} - \delta^{j}_{k}x^{l} - \delta^{i}_{k}q^{ij}x^{j}\delta^{l}_{i},\\
&= q^{lj}\delta^{l}_{k}x^{j} + q^{lj}\delta^{j}_{k}q^{jl}x^{l} - \delta^{j}_{k}x^{l} - \delta^{i}_{k}q^{ij}x^{j}\delta^{l}_{i},\\
&= q^{lj}\delta^{l}_{k}x^{j} + \underbrace{\delta^{j}_{k}x^{l}}_{for\;q^{lj}q^{jl} = 1}- \delta^{j}_{k}x^{l} - \delta^{l}_{k}q^{lj}x^{j} = 0.
\end{align*}
This shows that $I$ is T-derivative Consistent. We conclude that the factor algebra
\begin{equation*}
A= \bigslant{\mathbb{F}\langle x^{1}, \ldots , x^{n}\rangle}{\left( q^{ij}x^{i}x^{j} - x^{j}x^{i}\vert i<j\right) }.
\end{equation*}
has a FOCDC with the commutation rules (\ref{eq11}). In other words the relations (\ref{eq11}) defines a FOCDC on the factor algebra $A$.\\
For this algebra to be optimal, by the result (\ref{re1}), it is enough to verify that any homogeneous element of positive degree which has all partial derivatives zero is equal to zero in $A$ (i.e. zero modulo $I$).\\
We compute
\begin{equation*}
\partial_{k}[(x^{j})^{m}] = \partial_{k}[x^{j}\centerdot (x^{j})^{m-1}] = \partial_{k}x^{j}\centerdot (x^{j})^{m-1} + T^{i}_{k}(x^{j})\partial_{i}(x^{j})^{m-1}.
\end{equation*}
Using $T^{i}_{k}(x^{j}) = \delta^{i}_{k}q^{ij}x^{j}$, we write this as
\begin{equation}
\partial_{k}[(x^{j})^{m}] = \delta^{j}_{k}(x^{j})^{m-1} + \delta^{i}_{k}q^{ij}x^{j}\partial_{i}(x^{j})^{m-1}. \label{eq21}
\end{equation}
Similarly,
\begin{align*}
\partial_{i}(x^{j})^{m-1} &= \partial_{i}[x^{j}\centerdot (x^{j})^{m-2}] = \partial_{i}x^{j}\centerdot (x^{j})^{m-2} + \underbrace{T^{l}_{i}(x^{j})}_{=\delta^{l}_{i}q^{lj}x^{j}}\partial_{l}(x^{j})^{m-2},\\
&= \delta^{j}_{l}(x^{j})^{m-2} + \delta^{l}_{i}q^{lj}x^{j}\partial_{l}(x^{j})^{m-2}.
\end{align*}
Substituting this expression for $\partial_{i}(x^{j})^{m-1}$ in the relation (\ref{eq21}), we obtain
\begin{align}
\nonumber\partial_{k}[(x^{j})^{m}] &= \delta^{j}_{k}(x^{j})^{m-1} + \delta^{i}_{k}q^{ij}x^{j}\left[ \delta^{j}_{i}(x^{j})^{m-2} + \delta^{l}_{i}q^{lj}x^{j}\partial_{l}(x^{j})^{m-2}\right],\\
\nonumber &= \delta^{j}_{k}(x^{j})^{m-1} + \delta^{j}_{k}q^{jj}(x^{j})^{m-1} + \delta^{j}_{k}\delta^{i}_{j}\delta^{l}_{i}q^{ij}q^{lj}(x^{j})^{2}\partial_{l}(x^{j})^{m-2},\\
\nonumber &= \delta^{j}_{k}(x^{j})^{m-1} + \delta^{j}_{k}q^{jj}(x^{j})^{m-1} + \delta^{j}_{k}(q^{jj})^{2}(x^{j})^{2}\partial_{l}(x^{j})^{m-2},\\
\nonumber &= \ldots = \delta^{j}_{k}\left[ 1+ q^{jj} + (q^{jj})^{2} + \ldots + (q^{jj})^{m-1}\right](x^{j})^{m-1},\\
\therefore\quad\partial_{k}(x^{j})^{m} &= \delta^{j}_{k}\left[ 1+ q^{jj} + (q^{jj})^{2} + \ldots + (q^{jj})^{m-1}\right](x^{j})^{m-1}.   \label{eq13}
\end{align}
Next, we notice that an arbitrary element of the algebra $A$ has a unique presentation in the form
\begin{equation}
f = \sum\alpha_{i}(x^{1})^{i_{1}}(x^{2})^{i_{2}}\ldots (x^{n})^{i_{n}}.
\end{equation}
where $i\equiv (i_{1}, i_{2}, \ldots , i_{n})$ is a multi-index. Thus, by the relation (\ref{eq13}) we can write,
\begin{equation}
\partial_{k}f = \sum\alpha_{i}(x^{1})^{i_{1}}(x^{2})^{i_{2}}\ldots \partial_{k}[(x^{k})^{i_{k}}]\ldots (x^{n})^{i_{n}} = \sum (q^{kk})^{[i_{k}]}\alpha_{i}(x^{1})^{i_{1}}(x^{2})^{i_{2}}\ldots (x^{k})^{i_{k} -1}\ldots (x^{n})^{i_{n}}. \label{eq22}
\end{equation}
where by definition $(q^{kk})^{[i_{k}]} = 1+ q^{kk} + (q^{kk})^{2} + \ldots + (q^{kk})^{i_{k} -1}$.\\
Now, suppose $m$ is the least positive integer such that non of the scalars $(q^{kk})^{[m]},\; k= 1, \ldots ,n$ is zero. Then the relation (\ref{eq21}) implies $\partial_{k}f = 0\Longleftrightarrow \alpha_{i} = 0\Longleftrightarrow f=0$ in $A$. This implies that $A$ is the optimal algebra.
\item If $(q^{ii})^{[m_{i}]} = 0,\; 1\leq i\leq s$, then by the relation (\ref{eq13}) we have $\partial_{k}[(x^{i})^{m_{i}}] =0$ and also $\partial_{k}[T\lbrace (x^{j})^{mj}\rbrace] =0$, because $T^{i}_{k}(x^{j}) = \delta^{i}_{k}q^{ij}x^{j}$ implies that $T\lbrace (x^{j})^{m_{j}}\rbrace$ has the form $R\centerdot (x^{j})^{m_{j}}$ where $R\in M_{2}(\mathbb{F})$. Therefore, we must have $(x^{j})^{m_{j}} = 0$ in the optimal algebra for $i=1, \ldots ,s$. Let us consider the algebra
\begin{equation*}
\mathcal{A} := \bigslant{\mathbb{F}\langle x^{1}, \ldots ,x^{n}\rangle}{\left( q^{ij}x^{i}x^{j} = x^{j}x^{i},\;i<j,\; (x^{i})^{m_{i}},\; i=1, \ldots ,s\right) }
\end{equation*}
It follows that every element of this algebra has a unique presentation
\begin{equation*}
f = \sum\alpha_{i}(x^{1})^{i_{1}}(x^{2})^{i_{2}}\ldots (x^{s})^{i_{s}},\qquad i_{1}<m_{1},\; i_{2}<m_{2},\ldots ,i_{s}<m_{s}.
\end{equation*}
The formulas (\ref{eq13}) and (\ref{eq22}), which are still valid in $\mathcal{A}$ are zero in $\mathcal{A}$, imply that if all partial derivatives of an element in $\mathcal{A}$ are zero in $\mathcal{A}$, then this element is zero in $\mathcal{A}$. Hence $\mathcal{A}$ is the optimal algebra in this case.
\end{enumerate} 
\end{proof-non}
\begin{exmp}
Let $T= 0$, i.e. $x^{i}\centerdot dx^{j} =0$. Then $d$ of $\mathcal{F}$ is a homomorphism if right modules and the optimal algebra $A(T) = \mathcal{F}$.
\end{exmp}
\begin{proof-non}
By definition the cover differential (i.e. the free differential in $\mathcal{F}$) is a linear (i.e. $\mathbb{F}$-linear) mapping $d: \mathcal{F}\longrightarrow \mathcal{F}\centerdot d(\mathcal{F})\centerdot \mathcal{F}$ given by
\begin{equation}
x^{j}\centerdot dx^{i} = dx^{k}\centerdot T^{j}_{k}(x^{j}). \label{eq23}
\end{equation} 
If $T: \mathcal{F}\longrightarrow M_{n}(\mathcal{F})$ is the zero homomorphism, it follows from the relation (\ref{eq23}) that $x^{j}\centerdot dx^{i} =0,\; \forall i,j = 1, \ldots ,n$. Hence, $\mathcal{F}\centerdot d(\mathcal{F}) =0$; and in this case the cover differential is a linear mapping of $\mathcal{F}$-modules, $d: \mathcal{F}\longrightarrow d(\mathcal{F})\centerdot\mathcal{F}$. Moreover,
\begin{equation}
\forall f, g\in\mathcal{F}:\quad\partial_{k}(fg) = \partial_{k}(f)\centerdot g + T^{i}_{k}(f)\centerdot \partial_{i}g = \partial_{k}(f)\centerdot g,\quad for\;T =0,
\end{equation}
which shows that $\partial_{k}: \mathcal{F}\longrightarrow \mathcal{F}$ is a homomorphism of right $\mathcal{F}$-modules. It follows that
\begin{equation*}
\forall f, g\in\mathcal{F}:\quad d(fg) = dx^{i}\centerdot \partial_{i}(fg) = dx^{i}\centerdot (\partial_{i}f)g = (dx^{i}\centerdot \partial_{i}f)\centerdot g = df\centerdot g.
\end{equation*}
which shows that $d: \mathcal{F}\longrightarrow d(\mathcal{F})\centerdot \mathcal{F}$ is a homomorphism of right $\mathcal{F}$-modules.\\
Finally any ideal $I\vartriangleleft\mathcal{F}$ is sent to zero under $T= 0$, so it is T-Consistent. Let $f\in\mathcal{F}$ as a right $\mathcal{F}$-module. Then
\begin{equation*}
f = x^{1}f_{1} + \ldots + x^{n}f_{n},\;f_{i}\in\mathcal{F},
\end{equation*}
as an algebra. Then 
\begin{equation*}
\partial_{k}(f) = \partial_{k}(x^{i}f_{i}) = \delta^{i}_{k}f_{i} = f_{k},
\end{equation*}
hence 
\begin{equation*}
\partial_{k}(f) =0\;for\;all\;k=1, \ldots ,n\Longleftrightarrow f_{k} =0\;for\;all\;k=1, \ldots ,n\Longrightarrow f=0.
\end{equation*}
Therefore, the ideal $I\vartriangleleft\mathcal{F}$ cannot have a non-zero algebra is $\mathcal{F}$.
\end{proof-non}
\begin{exmp}\label{ex44}
Let $x^{i}\centerdot dx^{j} = -dx^{i}\centerdot x^{j},\; i,j= 1, \ldots ,n$. We shall prove that the optimal algebra is the smallest possible algebra generated by the space $V= lin_{\mathbb{F}}\lbrace x^{1}, \ldots ,x^{n}\rbrace$; that is
\begin{equation}
A(T) = \bigslant{\mathbb{F}\langle x^{1}, \ldots , x^{n}\rangle}{(x^{i}x^{j}\vert i,j= 1, \ldots ,n)}.
\end{equation}
\end{exmp}
\begin{proof-non}
As before, because we know that $\mathcal{F} = \mathbb{F}\langle x^{1}, \ldots , x^{n}\rangle$ accept any differential calculus, we consider
\begin{equation}
x^{i}\centerdot dx^{j} = -dx^{i}\centerdot x^{j},\qquad i,j = 1, \ldots ,n. \label{eq15}
\end{equation}
as a FOCDC on $\mathcal{F}$ and work with the free differential map $d$ of $\mathcal{F}$ (i.e. the cover differential). We immediately obtain, using the relation (\ref{eq15}),
\begin{equation*}
d(x^{i}x^{j}) = dx^{i}\centerdot x^{j} + x^{i}\centerdot dx^{j} =0.
\end{equation*}
Further we know that the space of all quadratic forms is T-invariant, since $T$ is a homogeneous homomorphism. Therefore, we conclude, by the result (\ref{re1}), that in the optimal algebra we must have $x^{i}x^{j} = 0,\; i,j= 1, \ldots ,n$.
\end{proof-non}
\begin{exmp}
Let 
\begin{equation}
\left\lbrace\begin{array}{rl}x^{1}\centerdot dx^{1} &= dx^{1}\centerdot (\alpha_{2}x^{2} + \ldots + \alpha_{n}x^{n}),\quad \alpha_{i}\in\mathbb{F}\\
x^{i}\centerdot dx^{j} &= -dx^{j}\centerdot x^{i},\quad if\; i\neq 1\;or\;j\neq 1
\end{array}\right. \label{eq16}
\end{equation}
The optimal algebra is almost isomorphic to the algebra of polynomial in 1-variable. More precisely, 
\begin{equation*}
A(T) = \bigslant{\mathbb{F}\langle x^{1}, \ldots , x^{n}\rangle}{(x^{i}x^{j}\vert i\neq 1,\; j\neq 1)}
\end{equation*}
\end{exmp}
\begin{proof-non}
From the given commutation rules (\ref{eq16}), we immediately obtain
\begin{equation}
\left\lbrace\begin{array}{rl}T^{1}_{k}(x^{1}) &= \delta^{1}_{k}(\alpha_{2}x^{2} + \ldots + \alpha_{n}x^{n})\\
T^{j}_{k}(x^{i}) &= -\delta^{i}_{k}x^{j},\quad i\neq 1\;or\; j\cong 1
\end{array}\right. \label{eq17}
\end{equation}
Let $I\vartriangleleft\mathcal{F} = \mathbb{F}\langle x^{1}, \ldots , x^{n}\rangle$ be an ideal generated by
\begin{equation*}
x^{i}x^{j},\qquad i,j = 2, 3,\ldots , n,\quad i\neq 1,\;j\neq 1,
\end{equation*}
and let
\begin{equation*}
A := \bigslant{\mathcal{F}}{I}.
\end{equation*}
Because $\mathcal{F}$ is free, then it accepts the relations (\ref{eq16}) as a FOCDC on it and we may use the corresponding $d$ and $\partial$ to compute:
\begin{equation*}
\partial_{k}(x^{i}x^{j}) = \delta^{i}_{k}x^{j} + T^{s}_{k}(x^{i})\partial_{s}x^{j} = \delta^{i}_{k}x^{j} + T^{s}_{k}(x^{i})\delta^{j}_{s} = \delta^{i}_{k}x^{j} + T^{j}_{k}(x^{i}) =0\;if\; i\neq 1\;or\;j\neq 1.
\end{equation*}
Moreover, since $T$ is an algebra isomorphism,
\begin{equation*}
T^{m}_{s}(x^{i}x^{j}) = T^{l}_{s}(x^{i})T^{m}_{l}(x^{j}).
\end{equation*}
for all $i,j= 1, \ldots ,n$. It follows by the relation (\ref{eq17}) that for all values $i,j = 1, \ldots ,n\quad T^{m}_{s} (x^{i}x^{j}) = 0\; mod(I)$. This proves that $I$ is a T-Consistent (or T-invariant) ideal.\\
Next, any element of the factor algebra has a unique presentation of the form
\begin{equation*}
 f= \gamma_{k}x^{k} + \beta_{2}(x^{1})^{2} + \ldots + \beta_{n}(x^{1})^{N}.
\end{equation*}
We compute 
\begin{equation*}
\partial_{k}\left[ (x^{1})^{N}\right] = \partial_{k}\left[ x^{1}\centerdot (x^{1})^{N-1}\right] = \delta^{1}_{k}(x^{1})^{N-1} + T^{s}_{k}(x^{1})\partial_{s}[(x^{1})^{N-1}].  
\end{equation*}
If $k=1$, we have
\begin{equation*}
\partial_{1}[(x^{1})^{N}] = (x^{1})^{N-1} + T^{s}_{1}(x^{1})\partial_{s}[(x^{1})^{N-1}].
\end{equation*}
Now, by the relations (\ref{eq17}),
\begin{align*}
T^{s}_{1}(x^{1}) &= \left\lbrace\begin{array}{rl}(\alpha_{2}x^{2} + \ldots + \alpha_{n}x^{n}),\qquad &if\;s=1\\
-x^{1},\qquad &if\;s\neq 1
\end{array}\right. \\
\therefore\qquad \partial_{1}[(x^{1})^{N}] &= (x^{1})^{N-1} + (\alpha_{2}x^{2} + \ldots + \alpha_{n}x^{n})\centerdot\partial_{1}[(x^{1})^{N-1}] - \sum_{s\geq 2}x^{s}\partial_{s}[(x^{1})^{N-1}],\\
&= \left\lbrace\begin{array}{rl}x^{1} + (\alpha_{2}x^{2} + \ldots + \alpha_{n}x^{n}),\qquad &if\;N=2 \\
(x^{1})^{N-1},\qquad &if\; N\geq 3
\end{array}\right.
\end{align*}
Therefore, if $df = 0$ in $A$, then $\gamma_{k} = \partial_{k}(f) =0$ holds for $k\geq 2$; and
\begin{align*}
\partial_{1}f &= \gamma_{1} + \beta x^{1} + \beta_{2}(\alpha_{2}x^{2} + \ldots + \alpha_{n}x^{n}) + \beta_{3}(x^{1})^{2} + \ldots + \beta_{N}(x^{1})^{N-1} =0,\\
&\Longrightarrow \gamma_{1} = \beta_{2} = \ldots = \beta_{N} =0.
\end{align*}
This implies $f=0$. Hence $A(T) = A$.
\end{proof-non}
\begin{exmp}
Homogeneous commutation rules in two variables with commutation optimal algebras. We will here describe all homogeneous commutation rules in two variables with a commutative optimal algebra. In this case the ideal $I(T)$ is homogeneous, and commutativity of the optimal algebra is equivalent to
\begin{equation*}
[x^{1}, x^{2}] := x^{1}x^{2} - x^{2}x^{1}\in I_{2}.
\end{equation*}
where $I_{2}$ is the second homogeneous component of $I(T)$. \vspace{.5cm}\\
In what follows, we will call the commutation rule and the corresponding optimal algebra regular if the space $I_{2}$ is one dimensional, that is , it is generated by the commutator over $\mathcal{F}$. In the oposite case, we will call the commutation rules and the corresponding optimal algebra irregular. For example the optimal algebras and the corresponding calculi in the example (\ref{ex11}) and (\ref{ex44}) are irregular.\\
Clearly, if the optimal algebra is isomorphic to the algebra of polynomials in two variables, then the commutation rules are regular but not vice versa, i.e. a regular optimal algebra need not be isomorphic to $\mathbb{F}[x^{1}, x^{2}]$, the polynomial algebra in two commuting variables.
\begin{result}
Let $u, v_{1}, v_{2}, w\in V := lin_{\mathbb{F}}\lbrace x^{1}, x^{2}\rbrace$ and $\lambda ,\mu\in\mathbb{F}$. A homogeneous commutation rule with regular commutative optimal algebra belongs (up to the exchange of variables $x^{1}\longleftrightarrow x^{2}$) to one of the following four classes:
\begin{enumerate}[(I)]
\item $\left\lbrace\begin{array}{rl}x^{1}\centerdot dx^{1} &= dx^{1}\centerdot u+ dx^{2}\centerdot v_{1}\\
x^{1}\centerdot dx^{2} &= dx^{1}\centerdot w + dx^{2}\centerdot (\lambda v_{1} + x^{1})\\
x^{2}\centerdot dx^{1} &= dx^{1}\centerdot (w+ x^{2}) + dx^{2}\centerdot (\lambda v_{1})\\
x^{2}\centerdot dx^{2} &= dx^{1}\centerdot (\lambda w) + dx^{2}\centerdot (\lambda^{2}v_{1} - \lambda u +w + \lambda x^{1} + x^{2})
\end{array}\right.
$
\item $\left\lbrace\begin{array}{rl}x^{1}\centerdot dx^{1} &= dx^{1}\centerdot (x^{1} + \mu v_{1} + v_{2}) + dx^{2}\centerdot v_{1}\\
x^{1}\centerdot dx^{2} &= dx^{1}\centerdot (\lambda v_{1}) + dx^{2}\centerdot (v_{2} + x^{1})\\
x^{2}\centerdot dx^{1} &= dx^{1}\centerdot (\lambda v_{1} + x^{2}) + dx^{2}\centerdot v_{2}\\
x^{2}\centerdot dx^{2} &= dx^{1}\centerdot (\lambda v_{2}) + dx^{2}\centerdot (\lambda v_{1} - \mu v_{2} + x^{2})
\end{array}\right.
$
\item $\left\lbrace\begin{array}{rl}x^{1}\centerdot dx^{1} &= dx^{1}\centerdot u\\
x^{1}\centerdot dx^{2} &= dx^{2}\centerdot x^{1}\\
x^{2}\centerdot dx^{1} &= dx^{1}\centerdot x^{2}\\
x^{2}\centerdot dx^{2} &= dx^{2}\centerdot v_{1}
\end{array}\right.
$
\item $\left\lbrace\begin{array}{rl}x^{1}\centerdot dx^{1} &= dx^{1}\centerdot u\\
x^{1}\centerdot dx^{2} &= dx^{2}\centerdot u\\
x^{2}\centerdot dx^{1} &= dx^{1}\centerdot x^{2} + dx^{2}\centerdot (u-x^{1})\\
x^{2}\centerdot dx^{2} &= dx^{1}\centerdot w + dx^{2}\centerdot v_{1}
\end{array}\right.
$
\end{enumerate}
\end{result}
\begin{proof-non}
First of all we shall prove that each commutation rule has a commutative optimal algebra. For this purpose, it is enough to prove that the ideal generated by the commutator $[x^{1}, x^{2}] := x^{1}x^{2} - x^{2}x^{1}$, is Consistent.
\begin{enumerate}[\textbf{Case}~(I)]
\item We can easily read of the entries of the matrices $T(x^{1}), T(x^{2})$ from the commutation rules given for this case:\label{case1}
\begin{align}
& \left\lbrace\begin{array}{rl}T(x^{1}) &= \left(\begin{array}{cc}u& w\\
v_{1} & \lambda v_{1} + x^{1}
\end{array}\right)
 \\
T(x^{2}) &= \left(\begin{array}{cc}w+ x^{2}& \lambda w\\
\lambda v_{1} & \lambda^{2}v_{1} - \lambda u + w + \lambda x^{1} + x^{2}
\end{array}\right)
\end{array}\right. \label{eqc1}\\
\nonumber &\Longrightarrow T(x^{2}) = \lambda T(x^{1}) + (w + x^{2} - \lambda u)1_{2},\qquad 1_{2} = \left(\begin{array}{cc}1& 0\\
0 & 1
\end{array}\right)
\end{align}
Then it is easily verified that 
\begin{align*}
T([x^{1}, x^{2}]) &= [T(x^{1}), T(x^{2})] = [T(x^{1}), \lambda T(x^{1}) + (w + x^{2}-\lambda u)1_{2}]\\
&= \lambda [T(x^{1}), T(x^{1})] + (w+ x^{2} -\lambda u)[T(x^{1}), 1_{2}] =0\\
&\Longrightarrow T^{i}_{k}(I)\subseteq I.
\end{align*}
where $I= (x^{1}x^{2} - x^{2}x^{1})\vartriangleleft\mathcal{F} = \mathbb{F}\langle x^{1}, x^{2}\rangle$.\\
For the T-derivative invariance (or Consistency) we compute
\begin{align*}
\partial_{1}(x^{1}x^{2} - x^{2}x^{1}) &= x^{2} + T^{k}_{1}(x^{1})\partial_{k}(x^{2}) - T^{k}_{1}(x^{2})\partial_{k}(x^{1}) = x^{2} + T^{2}_{1}(x^{1})- T^{1}_{1}(x^{2}),\\
&= x^{2} + w - (w+ x^{2}) =0.
\end{align*}
where the relation (\ref{eqc1}) has been used. (Notice that in the matrix $(T^{i}_{j}),\; i$ is the column and $j$ is the row index).\\
Similarly,
\begin{align*}
\partial_{2}(x^{1}x^{2} - x^{2}x^{1}) &= T^{k}_{2}(x^{1})\partial_{k}x^{2} - x^{1} - T^{k}_{2}(x^{2})\partial_{k}(x^{1}) = T^{2}_{2}(x^{1}) - x^{1} - T^{1}_{2}(x^{2}),\\
&= \lambda v_{1} + x^{1} -x^{1} - \lambda v_{1} = 0.
\end{align*}
Therefore, the ideal $I= \left( [x^{1}, x^{2}]\right) $ is both T-Consistent and T-derivative Consistent, hence the algebra 
\begin{equation*}
A = \bigslant{\mathbb{F}\langle x^{1}, x^{2}\rangle}{(x^{1}x^{2} - x^{2}x^{1})}.
\end{equation*}
has a FOCDC given by the case (\ref{case1}). Because this ideal must be contained in the ideal $I(T)$ of the optimal algebra $A(T)$, we conclude that $A(T)$, the optimal algebra for this FOCDC is also commutative.
\item We directly read off the matrices $T(x^{1}), T(x^{2})$ from the given commutation rules and find \label{case2}
\begin{align*}
T(x^{1}) &= \left(\begin{array}{cc}x^{1} + \mu v_{1} + v_{2}& \lambda v_{1}\\
v_{1} & v_{2} + x^{1}
\end{array}\right) = 1_{2}x^{1} + \left(\begin{array}{cc}\mu v_{1} + v_{2}& \lambda v_{1}\\
v_{1} & v_{2}
\end{array}\right)\\
T(x^{2}) &= \left(\begin{array}{cc}\lambda v_{1} + x^{2}& \lambda v_{2}\\
v_{2} & \lambda v_{1} - \mu v_{2} + x^{2}
\end{array}\right) = 1_{2}x^{2} + \left(\begin{array}{cc}\lambda v_{1}& \lambda v_{2}\\
v_{2} & \lambda v_{1}-\mu v_{2}
\end{array}\right)
\end{align*}
It easily follows that
\begin{equation*}
T([x^{1}, x^{2}]) = [T(x^{1}), T(x^{2})] \equiv 0\;mod(I)\Longrightarrow T^{i}_{k}\subseteq I.
\end{equation*}
where $I= (x^{1}x^{2} - x^{2}x^{1})$; and this means that the ideal $I$ is T-consistent.\\
To check the T-derivative invariance we compute
\begin{align*}
\partial_{1}(x^{1}x^{2} - x^{2}x^{1}) &= \partial_{1}x^{1}\centerdot x^{2} + T^{k}_{1}(x^{1})\partial_{k}x^{2} - \partial_{1}x^{2}\centerdot x^{1} - T^{k}_{1}(x^{2})\partial_{k}x^{1} = x^{2} + T^{2}_{1}(x^{1}) - T^{1}_{1}(x^{2}),\\
&= x^{2} + \lambda v_{1} - \lambda v_{1} - x^{2} =0.\\
\partial_{2}(x^{1}x^{2} - x^{2}x^{1}) &= T^{k}_{2}(x^{1})\partial_{k}x^{2} - x^{1} - T^{k}_{2}(x^{2})\partial_{k}x^{1} = T^{2}_{2}(x^{1}) - x^{1} - T^{1}_{2}(x^{2}),\\
&= v_{2} + x^{1} - x^{1} - v_{2} =0.
\end{align*}
It follows that the commutation rules in the case (\ref{case2}) have a commutative optimal algebra.
\item We read the entries of $T(x^{1})$ and $T(x^{2})$ from the given commutation rules and obtain \label{case3}
\begin{equation*}
T(x^{1}) = \left(\begin{array}{cc}u& 0\\
0 & x^{1}
\end{array}\right), \qquad T(x^{2}) = \left(\begin{array}{cc}x^{2}& 0\\
0 & v_{1}
\end{array}\right).
\end{equation*}
Because $T(x^{1})T(x^{2}) = T(x^{2})T(x^{1})$, we conclude that $T^{i}_{k}(I)\subseteq I$; so $I= (x^{1}x^{2} - x^{2}x^{1})$ is T-Consistent. \\
To check consistency for the partial derivatives we compute 
\begin{align*}
\partial_{1}(x^{1}x^{2} - x^{2}x^{1}) &= x^{2} + T^{k}_{1}(x^{1})\partial_{k}x^{2} - \partial_{1}x^{2}\centerdot x^{1} - T^{k}_{1}(x^{2})\partial_{k}x^{1} = x^{2} + T^{2}_{1}(x^{1}) - T^{1}_{1}(x^{2}),\\
&= x^{2} +0 - x^{2} =0.\\
\partial_{2}(x^{1}x^{2} - x^{2}x^{1}) &= x^{2} + T^{k}_{1}(x^{1})\partial_{k}x^{2} - 0 - T^{k}_{1}(x^{2})\partial_{k}x^{1} = x^{2} + T^{2}_{1}(x^{1}) - T^{1}_{1}(x^{2}),\\
&= x^{2} + 0 - x^{2} =0.
\end{align*}
We conclude that the optimal algebra for this set of commutation rules is commutative because the ideal $I(T)$ must contain the ideal $I= (x^{1}x^{2} - x^{2}x^{1})$.
\item In this case we obtain \label{case4}
\begin{equation*}
T(x^{1}) = \left(\begin{array}{cc}u& 0\\
0 & u
\end{array}\right),\qquad T(x^{2}) = \left(\begin{array}{cc}x^{2}& w\\
u-x^{1} & v_{1}
\end{array}\right).
\end{equation*}
Because $T(x^{1})$ is a multiple of the identity matrix, one immediately concludes $[T(x^{1}), T(x^{2})] = 0$, i.e. 
\begin{equation*}
T([x^{1}, x^{2}]) = 0\Longrightarrow T^{i}_{k}(I)\subseteq I.
\end{equation*}
so $I$ is T-Consistent (or invariant). Moreover, we find
\begin{align*}
\partial_{1}(x^{1}x^{2} - x^{2}x^{1}) &= x^{2} + T^{k}_{1}(x^{1})\partial_{k}x^{2} - \partial_{1}x^{2}\centerdot x^{1} - T^{k}_{1}(x^{2})\centerdot \partial_{k}x^{1} = x^{2} + T^{2}_{1}(x^{1}) -0 - T^{1}_{1}(x^{2}),\\
&= x^{2} + 0 -0 - x^{2} =0.
\end{align*} 
Similarly,
\begin{equation*}
\partial_{2}(x^{1}x^{2} - x^{2}x^{1}) = T^{2}_{2}(x^{1}) - x^{1} - T^{1}_{2}(x^{2}) = u - x^{1} - (u - x^{1}) =0.
\end{equation*}
We conclude that the commutation relations in this class have a commutative optimal algebra for $I(T)$ contains $I$.\vspace{.5cm}\\
Conversely, let the commutation rule 
\begin{equation}
x^{i}\centerdot dx^{j} = dx^{k}\centerdot T^{j}_{k}(x^{i}).
\end{equation}
(We recall the notation 
\begin{equation}
T^{i\;j}_{\;\;k} := T^{j}_{k}(x^{i}).
\end{equation}
is the (j-k) $\equiv$ (column, row) entry of the matrix $T^{i} := T(x^{i})$) have commutation optimal algebra. This implies that 
\begin{align}
\nonumber 0 &= \partial_{k}(x^{i}x^{j} - x^{j}x^{i}) = \delta^{i}_{k}x^{j} + T^{i\;j}_{\;\;\;k} - \delta^{j}_{k}x^{i} - T^{j\;i}_{\;\;\;k},\\
&\Longrightarrow\left\lbrace\begin{array}{rl}T^{i\;j}_{\;\;k} &= T^{j\;i}_{\;\;k},\qquad if\;i\neq k\;and\;j\neq k\\
T^{i\;j}_{\;\;k} &= x^{j} + T^{j\;i}_{\;\;k},\qquad if\; j\neq i\;and\; k=j \label{eq5}
\end{array}\right. 
\end{align}
In case $n=2$, which is the case under consideration, the equation (\ref{eq5}) reduce to
\begin{equation}
T^{1\;2}_{\;\;\;2} = x^{1} + T^{2\;1}_{\;\;\;2},\qquad  T^{2\;1}_{\;\;\;1} = x^{2} + T^{1\;2}_{\;\;\;1}. \label{eq6}
\end{equation}
These are correspondence of T-derivative invariance. \\
Next, the T-Consistency (or T-invariance)
\begin{equation*}
T(x^{1}x^{2} - x^{2}x^{1}) =0.
\end{equation*}
implies that the matrices $T^{i} := T(x^{i}) =: T^{i\;s}_{\;\;k}$ and $T^{j} := T(x^{j}) =: T^{j\;s}_{\;\;k}$, commute in the ring of matrices over $A$: 
\begin{equation}
T^{i\;s}_{\;\;k}T^{j\;l}_{\;\;s} = T^{j\;s}_{\;\;k}T^{i\;l}_{\;\;s}. \label{eq7}
\end{equation}
Let us consider these equalities in detail. All $T^{i\;s}_{\;\;k}$ have degree one, so they are in the space $V:= lin_{\mathbb{F}}\lbrace x^{1}, x^{2}\rbrace$. It follows that relations (\ref{eq7}) have degree two and have belong to the second homogeneous component $I_{2}$ of the ideal $I(T)$. Let $A_{2} := \bigslant{\mathcal{F}}{I_{2}}$; which is a commutative algebra such that the relations (\ref{eq7}) are valid in it. As $A$ is regular, the space $I_{2}$ is generated by the commutator $(x^{1}x^{2} - x^{2}x^{1})$ and the algebra $A_{2} = \mathbb{F}[x^{1}, x^{2}]$ is the algebra of polynomials in two commuting variables.\vspace{.25cm}\\
If one of the matrices $T^{1} := T(x^{1}), T^{2} := T(x^{2})$ is scalar, then if necessary by renaming the variables, we can suppose $T(x^{1}) = \left(\begin{array}{cc}u& 0\\
0 & u
\end{array}\right)
$, and then relations (\ref{eq6}) yield
\begin{equation*}
\left\lbrace\begin{array}{rl}\left( T(x^{2})\right)^{1}_{1} &=: T^{2\;1}_{\;\;\;1} = x^{2}  \\
\left( T(x^{2})\right)^{1}_{2} &=: T^{2\;1}_{\;\;\;2} = u- x^{1} 
\end{array}\right.
\end{equation*}
This implies that the commutation rules belong to the series (\ref{case4}).\\
If both matrices $T(x^{1}), T(x^{2})$ are diagonal, then the relations (\ref{eq6}) immediately imply
\begin{equation*}
T(x^{1}) = \left(\begin{array}{cc}u& 0\\
0 & x^{1}
\end{array}\right),\qquad T(x^{2}) = \left(\begin{array}{cc}x^{2}& 0\\
0 & v_{1}
\end{array}\right)
\end{equation*}
and the commutation rules belong to the series (\ref{case3}).\vspace{.5cm}\\
Let us, therefore, suppose that no one of $T(x^{1}), T(x^{2})$ is scalar and one of them is not diagonal. We use the fact that in the algebra of $2\times 2$ matrices over the field of rational functions $K= \mathbb{F}(x^{1}, x^{2})$, the dimension of the centralizer of any $2\times 2$ scalar matrix over $\mathbb{F}$ is equal to $2$. This means that the centralizer of the matrix $T(x^{1})$ is generated by two matrices, namely $T(x^{1})$ and $1_{2}$. This implies that
\begin{equation*}
T(x^{2}) = g\centerdot T(x^{1}) + f\centerdot 1_{2},\quad g, f\in K.
\end{equation*}
It follows from this relation that
\begin{equation*}
T^{2\;2}_{\;\;\;1} = gT^{1\;2}_{\;\;\;1},\quad T^{2\;1}_{\;\;\;2} = gT^{1\;1}_{\;\;\;2}\Longrightarrow T^{2\;1}_{\;\;\;2}\centerdot T^{1\;2}_{\;\;\;2} = gT^{1\;1}_{\;\;\;2}.
\end{equation*}
Now, we use the fact that all entries of the matrices involved are linear combinations of the variables, so
\begin{equation*}
T^{2\;1}_{\;\;\;2} = \lambda T^{1\;1}_{\;\;\;2},\quad T^{2\;2}_{\;\;\;1} = \lambda T^{1\;2}_{\;\;\;1},
\end{equation*}
or
\begin{equation*}
T^{1\;2}_{\;\;\;1} = \lambda T^{1\;1}_{\;\;\;2},\quad T^{2\;2}_{\;\;\;1} = \lambda T^{2\;1}_{\;\;\;2}.
\end{equation*}
where $\lambda\in\mathbb{F}$ or $\lambda = \infty$. The last case $\lambda = \infty$ means $T^{1\;1}_{\;\;\;2} = T^{1\;2}_{\;\;\;1} =0$ or $T^{1\;1}_{\;\;\;2} = T^{2\;1}_{\;\;\;2} = 0$. These cases reduce to the case $\lambda = 0$ by changing variables $x^{1}\longleftrightarrow x^{2}$.\\
If $T^{2\;1}_{\;\;\;2} = \lambda T^{1\;1}_{\;\;\;2}, T^{2\;2}_{\;\;\;1} = \lambda T^{1\;2}_{\;\;\;1} -1$, then denoting $T^{1\;1}_{\;\;\;1} = u$ and $T^{1\;2}_{\;\;\;1} =w$, we obtain
\begin{equation*}
T^{2\;1}_{\;\;\;2} = \lambda v_{1},\qquad T^{2\;2}_{\;\;\;1} = \lambda w,
\end{equation*}
and therefore, $f= x^{2} + w -\lambda u$. This means that 
\begin{equation*}
T^{2\;2}_{\;\;\;2} = \lambda T^{1\;2}_{\;\;\;2} + f = \lambda x^{1} + \lambda^{2}v_{1} + x^{2} + w + \lambda u.
\end{equation*}
and this implies that the matrices $T(x^{1})$ and $T(x^{2})$ has the form like series (\ref{case1}).\vspace{.5cm}\\
If $T^{1\;2}_{\;\;\;1} = \lambda T^{1\;1}_{\;\;\;2},\; T^{2\;2}_{\;\;\;1} = \lambda T^{2\;1}_{\;\;\;2} $, then by denoting $T^{1\;1}_{\;\;\;2} = v_{1}$ and $T^{2\;1}_{\;\;\;2} = v_{2}$, we obtain
\begin{align*}
g &= \frac{v_{2}}{v_{1}},\; f = T^{2\;1}_{\;\;\;1} - \frac{v_{2}}{v_{1}}T^{1\;1}_{\;\;\;1} = T^{2\;2}_{\;\;\;2} - \frac{v_{2}}{v_{1}}T^{1\;2}_{\;\;\;2},\\
&\Longrightarrow v_{2}\left( T^{1\;1}_{\;\;\;1} - T^{1\;2}_{\;\;\;2}\right) = v_{1}\left( T^{2\;1}_{\;\;\;1} - T^{2\;2}_{\;\;\;2}\right).  
\end{align*}
Because all the factors in this relation are linear combinations of the variables, we conclude
\begin{equation*}
T^{1\;1}_{\;\;\;1} - T^{1\;2}_{\;\;\;2} = \mu v_{1},\quad T^{2\;1}_{\;\;\;1} - T^{2\;2}_{\;\;\;2} = \mu v_{2}.
\end{equation*} 
where $\mu \in\mathbb{F}$. The relations (\ref{eq6}) take the following form in this case
\begin{equation*}
T^{1\;2}_{\;\;\;2} = x^{1} + v_{2},\qquad T^{2\;1}_{\;\;\;1} = x^{2} + \lambda v_{1}.
\end{equation*}
which imply that $T^{1\;1}_{\;\;\;1} = x^{1} + v_{2} + \mu v_{1}$ and $T^{2\;2}_{\;\;\;2} = x^{2} + \lambda v_{1} - \mu v_{2}$. This gives the case (\ref{case2}).
\end{enumerate}
\end{proof-non}
\end{exmp}
It is an open problem to determine the optimal algebra for the commutation rules described above. We have not claimed that the optimal algebras in two variables is
\begin{equation*}
\bigslant{\mathbb{F}\langle x^{1}, x^{2}\rangle}{I_{2}} = \mathbb{F}[x^{1}, x^{2}].
\end{equation*}
\bibliographystyle{unsrt}  
\bibliography{references}  


\end{document}